\documentclass[sigconf, authorversion]{acmart}

\AtBeginDocument{%
  }

\usepackage{balance}       %
\usepackage{graphics}      %
\usepackage{hyperref}
\usepackage{color}
\usepackage{booktabs}
\usepackage{textcomp}
\usepackage{enumerate}
\usepackage{xcolor}
\usepackage{lipsum}%
\usepackage{makecell}
\usepackage{multicol}
\usepackage{multirow}
\usepackage{array}
\usepackage{verbatimbox}
\usepackage{enumitem}
\usepackage{amsmath}
\usepackage{stfloats}
\usepackage{amsthm}
\usepackage{listings}
\usepackage{caption} 
\usepackage{xspace}
\usepackage[linesnumbered]{algorithm2e}
\usepackage{algpseudocode}
\usepackage{tabularx}
\usepackage{arydshln}
\usepackage[bottom]{footmisc}
\usepackage{stackengine}
\usepackage{placeins}
\usepackage{graphicx}
\usepackage{subcaption} %
\usepackage{mdframed}
\usepackage{pifont}

\newcommand*{\eg}{\textit{e.g.},\xspace}
\newcommand*{\ie}{\textit{i.e.},\xspace}
\newcommand*{\vs}{\textit{vs.}\xspace}

\newcolumntype{L}[1]{>{\raggedright\let\newline\\\arraybackslash\hspace{0pt}}m{#1}}
\newcolumntype{C}[1]{>{\centering\let\newline\\\arraybackslash\hspace{0pt}}m{#1}}
\newcolumntype{R}[1]{>{\raggedleft\let\newline\\\arraybackslash\hspace{0pt}}m{#1}}

\makeatletter
\def\thickhline{%
  \noalign{\ifnum0=`}\fi\hrule \@height \thickarrayrulewidth \futurelet
   \reserved@a\@xthickhline}
\def\@xthickhline{\ifx\reserved@a\thickhline
               \vskip\doublerulesep
               \vskip-\thickarrayrulewidth
             \fi
      \ifnum0=`{\fi}}
\makeatother

\makeatletter
\def\thickhlinespace{%
  \addlinespace[1ex]
  \noalign{\ifnum0=`}\fi\hrule \@height \thickarrayrulewidth \futurelet
   \reserved@a\@xthickhline
   \addlinespace[1ex]
   }
\def\@xthickhlinespace{\ifx\reserved@a\thickhline
               \vskip\doublerulesep
               \vskip-\thickarrayrulewidth
             \fi
      \ifnum0=`{\fi}}
\makeatother

\newlength{\thickarrayrulewidth}
\newlength\Origarrayrulewidth

\algnewcommand{\IfThenElse}[3]{%
  \State \algorithmicif\ #1\ \algorithmicthen\ #2\ \algorithmicelse\ #3}

\newcommand{\systemName}{MIND\xspace}
\newcommand{\baselineName}{FACT\xspace}
\newcommand{\z}{\textit{Z}}
\newcommand{\p}{\textit{p}}
\newcommand{\eff}{$\text{\textit{r}}_\text{\textit{b}}$}
\newcommand{\M}[2]{$\text{\textit{M}}_\text{\textit{#1}}$=#2}

\definecolor{downredcolor}{HTML}{e31a1c}
\definecolor{upgreencolor}{HTML}{33a02c}

\definecolor{DarkGreen}{HTML}{5DAC81}

\definecolor{revisionColor}{HTML}{025DF4}

\definecolor{maroon}{cmyk}{0, 0.87, 0.68, 0.32}
\definecolor{halfgray}{gray}{0.55}
\definecolor{ipython_frame}{RGB}{207, 207, 207}
\definecolor{ipython_bg}{RGB}{247, 247, 247}
\definecolor{ipython_red}{RGB}{186, 33, 33}
\definecolor{ipython_green}{RGB}{0, 128, 0}
\definecolor{ipython_cyan}{RGB}{64, 128, 128}
\definecolor{ipython_purple}{RGB}{170, 34, 255}
\lstdefinelanguage{Markdown}{
    basicstyle=\ttfamily\footnotesize,
}
\lstdefinelanguage{iPython}{
    morekeywords={access,and,break,class,continue,def,del,elif,else,except,exec,finally,for,from,global,if,import,in,is,lambda,not,or,pass,print,raise,return,try,while},%
    morekeywords=[2]{abs,all,any,basestring,bin,bool,bytearray,callable,chr,classmethod,cmp,compile,complex,delattr,dict,dir,divmod,enumerate,eval,execfile,file,filter,float,format,frozenset,getattr,globals,hasattr,hash,help,hex,id,input,int,isinstance,issubclass,iter,len,list,locals,long,map,max,memoryview,min,next,object,oct,open,ord,pow,property,range,raw_input,reduce,reload,repr,reversed,round,set,setattr,slice,sorted,staticmethod,str,sum,super,tuple,type,unichr,unicode,vars,xrange,zip,apply,buffer,coerce,intern},%
    sensitive=true,%
    morecomment=[l]\#,%
    morestring=[b]',%
    morestring=[b]",%
    morestring=[s]{'''}{'''},%
    morestring=[s]{"""}{"""},%
    morestring=[s]{r'}{'},%
    morestring=[s]{r"}{"},%
    morestring=[s]{r'''}{'''},%
    morestring=[s]{r"""}{"""},%
    morestring=[s]{u'}{'},%
    morestring=[s]{u"}{"},%
    morestring=[s]{u'''}{'''},%
    morestring=[s]{u"""}{"""},%
    literate=
    {á}{{\'a}}1 {é}{{\'e}}1 {í}{{\'i}}1 {ó}{{\'o}}1 {ú}{{\'u}}1
    {Á}{{\'A}}1 {É}{{\'E}}1 {Í}{{\'I}}1 {Ó}{{\'O}}1 {Ú}{{\'U}}1
    {à}{{\`a}}1 {è}{{\`e}}1 {ì}{{\`i}}1 {ò}{{\`o}}1 {ù}{{\`u}}1
    {À}{{\`A}}1 {È}{{\'E}}1 {Ì}{{\`I}}1 {Ò}{{\`O}}1 {Ù}{{\`U}}1
    {ä}{{\"a}}1 {ë}{{\"e}}1 {ï}{{\"i}}1 {ö}{{\"o}}1 {ü}{{\"u}}1
    {Ä}{{\"A}}1 {Ë}{{\"E}}1 {Ï}{{\"I}}1 {Ö}{{\"O}}1 {Ü}{{\"U}}1
    {â}{{\^a}}1 {ê}{{\^e}}1 {î}{{\^i}}1 {ô}{{\^o}}1 {û}{{\^u}}1
    {Â}{{\^A}}1 {Ê}{{\^E}}1 {Î}{{\^I}}1 {Ô}{{\^O}}1 {Û}{{\^U}}1
    {œ}{{\oe}}1 {Œ}{{\OE}}1 {æ}{{\ae}}1 {Æ}{{\AE}}1 {ß}{{\ss}}1
    {ç}{{\c c}}1 {Ç}{{\c C}}1 {ø}{{\o}}1 {å}{{\r a}}1 {Å}{{\r A}}1
    {€}{{\EUR}}1 {£}{{\pounds}}1
    {^}{{{\color{ipython_purple}\^{}}}}1
    {=}{{{\color{ipython_purple}=}}}1
    {+}{{{\color{ipython_purple}+}}}1
    {*}{{{\color{ipython_purple}$^\ast$}}}1
    {/}{{{\color{ipython_purple}/}}}1
    {+=}{{{+=}}}1
    {-=}{{{-=}}}1
    {*=}{{{$^\ast$=}}}1
    {/=}{{{/=}}}1,
    literate=
    *{-}{{{\color{ipython_purple}-}}}1
     {?}{{{\color{ipython_purple}?}}}1,
    identifierstyle=\color{black}\ttfamily,
    commentstyle=\color{ipython_cyan}\ttfamily,
    stringstyle=\color{ipython_red}\ttfamily,
    keepspaces=true,
    showspaces=false,
    showstringspaces=false,
    basicstyle=\ttfamily\footnotesize,
    keywordstyle=\color{ipython_green}\ttfamily,
}

\definecolor{lightOrange}{HTML}{FFBF86}

\definecolor{teaserGreen}{HTML}{255634}
\definecolor{teaserPurple}{HTML}{9661BC}
\definecolor{teaserBlue}{HTML}{4675AF}
\definecolor{teaserRed}{HTML}{B5404A}
\definecolor{teaserGray}{HTML}{5F5F5F}

\definecolor{captionOrange}{HTML}{FFA55B}
\definecolor{captionBlue}{HTML}{3D76DD}
\definecolor{captionGreen}{HTML}{008685}

\definecolor{summarizerColor}{HTML}{5E5E5E}
\definecolor{discovererColor}{HTML}{BF9515}
\definecolor{synthesizerColor}{HTML}{318BCA}
\definecolor{narratorColor}{HTML}{6053CC}

\definecolor{captionYellow}{HTML}{906E10}

\usepackage[table]{xcolor}
\usepackage{colortbl}
\usepackage{wrapfig}

\usepackage{acmart-taps}
\newcommand{\mathTextItalics}[1]{\text{\textit{#1}}}

\newcommand*\inline[1]{\protect\includegraphics[height=.9em]{#1}}
\newcommand{\inlinefig}[1]{\protect\raisebox{-.2em}{\inline{wsg/#1.pdf}}}

\usepackage{tikz}
\newcommand{\twiceinclude}[2][]{%
  \begin{tikzpicture}%
    \node at (0,0) {\includegraphics[#1]{#2.png}};%
    \node at (0,0) {\includegraphics[#1]{#2.pdf}};%
  \end{tikzpicture}%
}%

\usepackage{tikz}
\usepackage{xcolor}
\usepackage{xparse}

\definecolor{plotdata}{HTML}{1F77B4}       %
\definecolor{plotmean}{HTML}{D95F00}       %
\definecolor{plotmeanlabel}{HTML}{994D00}  %
\definecolor{plotticks}{HTML}{666666}      %
\newlength{\dotplotheight}
\newlength{\dotplotwidth}
\pgfmathsetlengthmacro{\plotwidthnum}{\dotplotwidth}
\pgfmathsetlengthmacro{\plotheightnum}{\dotplotheight}

\pgfmathsetmacro{\tickh}{0.2}    
\pgfmathsetmacro{\meanh}{0.5}     
\pgfmathsetmacro{\pointsize}{0.12}

\ExplSyntaxOn
\keys_define:nn { dotplot }
 {
  min  .tl_set:N  = \l_dotplot_min_tl,
  min  .initial:n = {0},
  max  .tl_set:N  = \l_dotplot_max_tl,
  max  .initial:n = {10},
  step .tl_set:N  = \l_dotplot_step_tl,
  step .initial:n = {1},
 }
\ExplSyntaxOff

\ExplSyntaxOn
\NewDocumentCommand{\dotplot}{ O{} m }
 {
  \group_begin:
  \keys_set:nn { dotplot } { #1 }

  \pgfmathsetmacro{\dotplotSum}{0}
  \pgfmathsetmacro{\dotplotCount}{0}
  \foreach \val in {#2} {
      \pgfmathparse{\dotplotSum + \val} \global\let\dotplotSum\pgfmathresult
      \pgfmathparse{\dotplotCount + 1} \global\let\dotplotCount\pgfmathresult
  }
  \pgfmathsetmacro{\mean}{ifthenelse(\dotplotCount>0,\dotplotSum/\dotplotCount,0)}

  \pgfmathsetmacro{\numticks}{int((\l_dotplot_max_tl - \l_dotplot_min_tl)/\l_dotplot_step_tl)}

\begin{tikzpicture}[x=\plotwidthnum/(\l_dotplot_max_tl - \l_dotplot_min_tl), y=\plotheightnum, baseline=-0.25\dotplotheight]
    \draw[plotticks] (\l_dotplot_min_tl,0) -- (\l_dotplot_max_tl,0);

    \pgfmathsetlengthmacro{\tickheightnum}{\tickh*\dotplotheight}
    \foreach \i in {0,...,\numticks} {
        \pgfmathsetmacro{\x}{\l_dotplot_min_tl + \i*\l_dotplot_step_tl}
        \ifnum\i=0
            \draw[plotticks, thick] (\x,-\tickheightnum) -- (\x,\tickheightnum);
            \node[overlay, font=\tiny, anchor=north, gray] at (\x+0.15,\tickheightnum-1) {\l_dotplot_min_tl};
        \else
            \ifnum\i=\numticks
                \draw[plotticks, thick] (\x,-\tickheightnum) -- (\x,\tickheightnum);
                \node[overlay, font=\tiny, anchor=north, gray] at (\x-0.15,\tickheightnum-1) {\l_dotplot_max_tl};
            \else
                \draw[plotticks] (\x,0) -- (\x,\tickheightnum);
            \fi
        \fi
    }

    \pgfmathsetlengthmacro{\radiusnum}{\pointsize*\dotplotheight}
    \def\jitter{0.15} %
    
    \foreach \val in {#2} {
      \pgfmathsetmacro{\yjitter}{\jitter*rand}
      \fill[plotdata, opacity = 0.6] (\val,\yjitter) circle[radius=\radiusnum];
    }

    \pgfmathsetlengthmacro{\meanheightnum}{\meanh*\dotplotheight}
    \draw[plotmean, line~width=1pt] (\mean,-\meanheightnum) -- (\mean,\meanheightnum);
    \node[overlay, font={\bfseries\tiny}, anchor=west, gray, plotmean] 
    at (\mean-0.1, \tickheightnum+2.2) 
    {\pgfmathprintnumber[fixed, precision=2]{\mean}};
  \end{tikzpicture}

  \group_end:
 }
\ExplSyntaxOff

\ExplSyntaxOn
\NewDocumentCommand{\pieplot}{ O{plotticks} m m }
 {
  \pgfmathsetmacro{\pieangle}{360*(#2)/(#3)}
  
  \begin{tikzpicture}[baseline=-0.8ex]
    \draw[#1] (0,0) circle (0.13cm);
    \fill[#1!60] (0,0) -- (0,0.13cm)
      arc [start~angle=90, end~angle={90-\pieangle}, radius=0.13cm] -- cycle;
  \end{tikzpicture}%
 }
\ExplSyntaxOff

\newcommand{\progressplot}[4][3.0]{%
  \begin{tikzpicture}[baseline=-0.25ex]
    \pgfmathsetmacro{\percentage}{(#2 - #3)/(#4 - #3)}
    \pgfmathsetmacro{\barwidth}{#1}
    \pgfmathsetmacro{\progresswidth}{\percentage * \barwidth}
    
    \fill[color=lightgray!50] (0, 0) rectangle +(\barwidth ex, 1.2ex);
    
    \fill[color=plotdata!75] (0, 0) rectangle +(\progresswidth ex, 1.2ex);

    \node[font=\bfseries\small, anchor=west, text=black, inner sep=1pt] 
      at (\barwidth ex + 0.3ex, 0.6ex) 
      {\pgfmathprintnumber[fixed, precision=2]{#2}};
  \end{tikzpicture}%
}

\acmSubmissionID{6290}

\begin{document}

\title{\systemName: Empowering Mental Health Clinicians with Multimodal Data Insights through a Narrative Dashboard}

\author{Ruishi Zou}
\authornote{Both authors contributed equally to this research.}
\affiliation{%
\institution{Columbia University}
\city{New York}
\state{New York}
\country{USA}}
\email{rz2794@cumc.columbia.edu}

\author{Shiyu Xu}
\authornotemark[1]
\affiliation{%
\institution{Columbia University}
\city{New York}
\state{New York}
\country{USA}}
\email{sx2430@cumc.columbia.edu}

\author{Margaret E. Morris}
\affiliation{%
  \institution{University of Washington}
  \city{Seattle}
  \state{Washington}
  \country{USA}}
\email{margiemm@uw.edu}

\author{Jihan Ryu}
\affiliation{%
  \institution{Hamilton-Madison House}
  \city{New York}
  \state{New York}
  \country{USA}}
\email{jihan.i.ryu@gmail.com}

\author{Timothy D. Becker}
\affiliation{%
  \institution{New York-Presbyterian Hospital/ Weill Cornell Medicine}
  \city{New York}
  \state{New York}
  \country{USA}}
\email{tdb2002@med.cornell.edu}

\author{Nicholas Allen}
\affiliation{%
  \institution{University of Oregon}
  \city{Eugene}
  \state{OR}
  \country{USA}}
\email{nallen3@uoregon.edu}

\author{Anne Marie Albano}
\affiliation{%
  \institution{Columbia University}
  \city{New York}
  \state{New York}
  \country{USA}}
\email{aa2289@cumc.columbia.edu}

\author{Randy Auerbach}
\affiliation{%
  \institution{Columbia University}
  \city{New York}
  \state{New York}
  \country{USA}}
\email{rpa2009@cumc.columbia.edu}

\author{Dan Adler}
\affiliation{%
  \institution{Cornell University}
  \city{New York}
  \state{New York}
  \country{USA}}
\email{daa243@cornell.edu}

\author{Varun Mishra}
\affiliation{%
  \institution{Northeastern University}
  \city{Boston}
  \state{MA}
  \country{USA}}
\email{v.mishra@northeastern.edu}

\author{Lace Padilla}
\affiliation{%
  \institution{Northeastern University}
  \city{Boston}
  \state{MA}
  \country{USA}}
\email{l.padilla@northeastern.edu}

\author{Dakuo Wang}
\affiliation{%
  \institution{Northeastern University}
  \city{Boston}
  \state{MA}
  \country{USA}}
\email{d.wang@northeastern.edu}

\author{Ryan Sultan}
\affiliation{%
  \institution{Columbia University}
  \city{New York}
  \state{New York}
  \country{USA}}
\email{rs3511@cumc.columbia.edu}

\author{Xuhai ``Orson'' Xu}
\affiliation{%
  \institution{Columbia University}
  \city{New York}
  \state{New York}
  \country{USA}}
\email{xx2489@cumc.columbia.edu}

\renewcommand{\shortauthors}{}
\renewcommand{\shorttitle}{}

\begin{abstract}

Advances in data collection enable the capture of rich patient-generated data: from passive sensing (e.g., wearables and smartphones) to active self-reports (e.g., cross-sectional surveys and ecological momentary assessments). Although prior research has demonstrated the utility of patient-generated data in mental healthcare, significant challenges remain in effectively presenting these data streams along with clinical data (e.g., clinical notes) for clinical decision-making. Through co-design sessions with five clinicians, we propose MIND, a large language model-powered dashboard designed to present clinically relevant multimodal data insights for mental healthcare. MIND presents multimodal insights through narrative text, complemented by charts communicating underlying data. Our user study (\textit{N}=16) demonstrates that clinicians perceive MIND as a significant improvement over baseline methods, reporting improved performance to reveal hidden and clinically relevant data insights (\p<.001) and support their decision-making (\p=.004). Grounded in the study results, we discuss future research opportunities to integrate data narratives in broader clinical practices.

\end{abstract}


\copyrightyear{2026}
\acmYear{2026}
\setcopyright{cc}
\setcctype{by-nc-nd}
\acmConference[CHI '26]{Proceedings of the 2026 CHI Conference on Human Factors in Computing Systems}{April 13--17, 2026}{Barcelona, Spain}
\acmBooktitle{Proceedings of the 2026 CHI Conference on Human Factors in Computing Systems (CHI '26), April 13--17, 2026, Barcelona, Spain}
\acmPrice{}
\acmDOI{10.1145/3772318.3790529}
\acmISBN{979-8-4007-2278-3/2026/04}



\maketitle

\begin{figure*}
    \centering
    \twiceinclude[width=.94\textwidth]{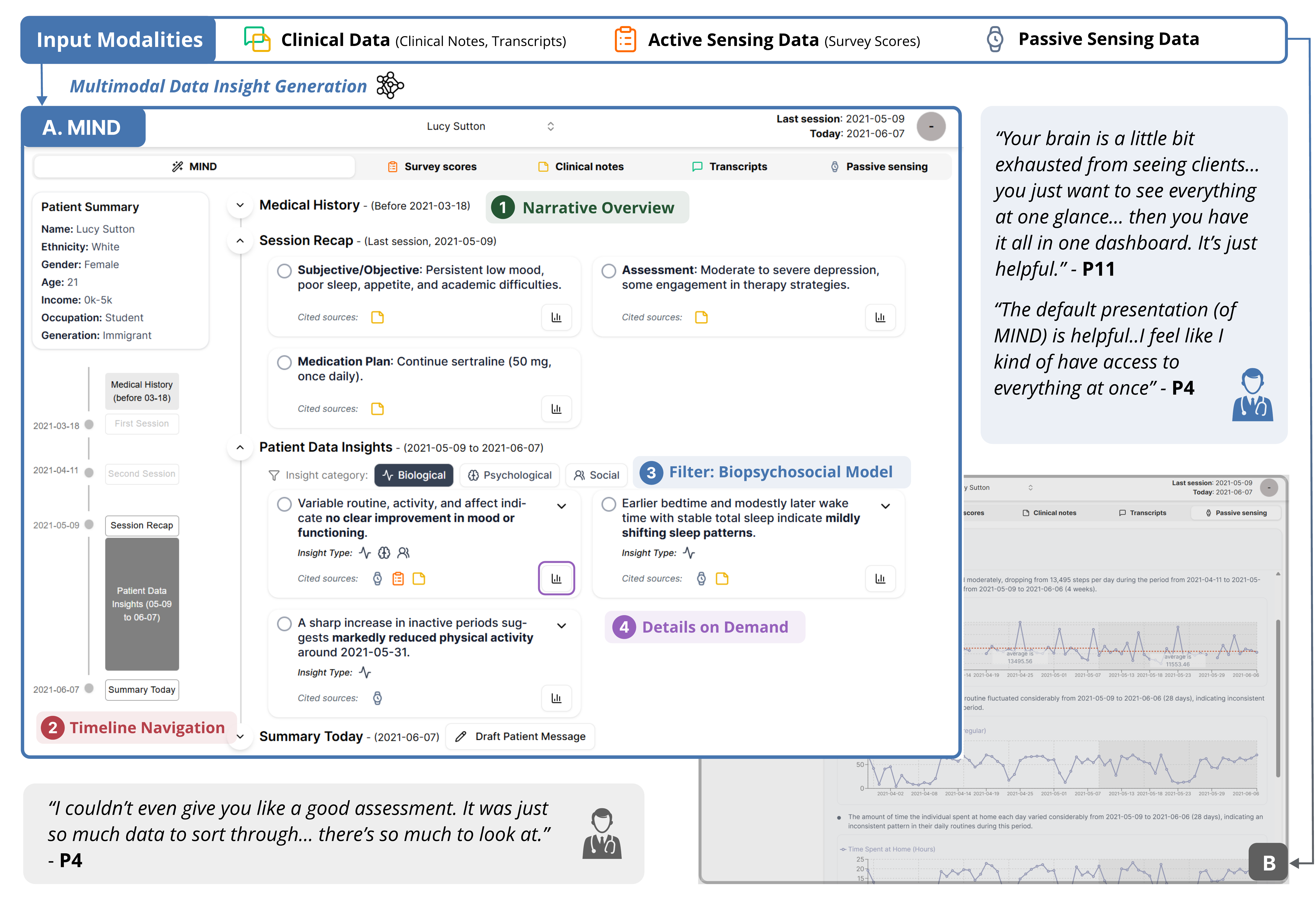}
    \caption{A comparison between the proposed narrative dashboard \textbf{\systemName (A)} with a data collection dashboard design \textbf{(B)}, showing the data from the same hypothetical patient. \systemName presents computationally generated multimodal insights in a \textcolor{teaserGreen}{narrative way \ding{182}}, allowing clinicians to better understand multimodal patient data. Clinicians can also easily \textcolor{teaserRed}{navigate through timeline \ding{183}}, \textcolor{teaserBlue}{filter by insight category \ding{184}}, and \textcolor{teaserPurple}{expand details on demand \ding{185}}.}
    \Description{The MIND dashboard for multimodal data insight generation. The system integrates clinical data (e.g., notes, transcripts), active sensing data (e.g., survey scores), and passive sensing data (e.g., wearable/device logs) into a single interface.
    Narrative Overview: presents medical history, session recap, and patient data insights with source citations to support clinical reasoning.
    Timeline Navigation: allows clinicians to move across historical and current patient information, from medical history to session recap to “Summary Today.”
    Filter by Biopsychosocial Model: enables clinicians to categorize and interpret insights across biological, psychological, and social domains.
    Details on Demand: interactive elements let clinicians expand summarized insights into raw underlying data.
    Quotes from study participants illustrate the value of MIND in reducing information overload and enabling efficient case review: “Your brain is a little bit exhausted… you just want to see everything at one glance” (P11), and “I feel like I kind of have access to everything at once” (P4). At the same time, clinicians noted challenges of volume and synthesis: “It was just so much data to sort through” (P4). Together, these features highlight how MIND balances concise narrative presentation with flexible access to underlying multimodal data.}
    \Description{
    The MIND dashboard for multimodal data insight generation. The system integrates clinical data (e.g., notes, transcripts), active sensing data (e.g., survey scores), and passive sensing data (e.g., wearable/device logs) into a single interface.
1. Narrative Overview: presents medical history, session recap, and patient data insights with source citations to support clinical reasoning.
2. Timeline Navigation: allows clinicians to move across historical and current patient information, from medical history to session recap to “Summary Today.”
3. Filter by Biopsychosocial Model: enables clinicians to categorize and interpret insights across biological, psychological, and social domains.
4. Details on Demand: interactive elements let clinicians expand summarized insights into raw underlying data.
Quotes from study participants illustrate the value of MIND in reducing information overload and enabling efficient case review: “Your brain is a little bit exhausted… you just want to see everything at one glance” (P11), and “I feel like I kind of have access to everything at once” (P4). At the same time, clinicians noted challenges of volume and synthesis: “It was just so much data to sort through” (P4). Together, these features highlight how MIND balances concise narrative presentation with flexible access to underlying multimodal data.
    }
    \label{fig:teaser}
\end{figure*}

\section{Introduction}
\label{sec:introduction}

Recent advances in technology have enabled patient-generated data to be captured at scale~\cite{demiris2019patient,heydari2025anatomy}. For instance, mobile phones and wearable devices can passively capture patients' biological, psychological, and behavioral markers longitudinally, while digital methods such as ecological momentary assessment (EMA) enable frequent self-reports of patients' states and contexts. While traditional clinical data sources such as electronic health records (EHRs) remain widely used, patient-generated data are increasingly being applied in mental health assessment and care~\cite{onnela2016harnessing, mohr2017personal, dinh-le2019wearableb,wang2025beyond}. Specifically, they afford clinicians an alternative and complementary perspective on patients' daily lives outside clinical settings--one that is difficult to obtain in current  practice.

Prior research has investigated three broad data modalities related to mental healthcare~\cite{adler2025designing}: 1) \textbf{clinical data} (\eg clinical notes, and transcripts), and two types of patient-generated data: 2) \textbf{active sensing data} (\eg survey scores), and 3) \textbf{passive sensing data} (\eg sleep, steps). 
One of the straightforward ways of leveraging these data modalities is to build machine learning (ML) models to ``detect/predict''~\cite{adler2024detectiona} mental health measures (\eg~stress~\cite{egilmez2017ustress}, depression~\cite{xu2019leveraginga}).
However, recent evidence suggests these models barely generalize across datasets~\cite{xu2023globema, adler2022machineb, meegahapola2023generalization}, limiting their real-world applicability.
Moreover, building detection/prediction ML models hardly provides clinicians with meaningful insights for treatment decision-making in real-world practices.
This leads to another thread of research that takes a more interpretive approach by viewing these data modalities as complementary perspectives to support more informed clinical decision-making~\cite{sivaraman2023ignorea}. However, without clinical abstraction or contextual explanation, the inherent complexity of multimodal data (\eg data format: textual clinical data \vs numerical sensing data; temporal frequency: daily passive sensing data \vs weekly active sensing data \vs monthly per-session clinical data) creates significant challenges for clinicians to synthesize and reference in practice~\cite{englhardt2024classificationc, birk2022digital, azodo2020opportunities}.

One prevalent solution to present multimodal data at scale is through clinical dashboards (\eg~\cite{wu2025cardioai}), a technique that promises rich data information ``at a glance''~\cite{few2006information}. Researchers have developed dashboards that accommodate clinical, active sensing, and passive sensing data in separate views (Fig.~\ref{fig:teaser}.B)~\cite{wu2025cardioai,yao2025more,li2025vital}--a design pattern coined as \textbf{data collection dashboard} by~\citet{bach2023dashboard}.
While they accommodate large volumes of information, data collection dashboards are designed with minimal curation of what and how the data is presented, leaving mental health clinicians to forage for relevant data and synthesize insights themselves. Given the time-sensitive nature of clinical practice, presenting multimodal data through data collection dashboards may impede their use in clinicians' daily workflow, undermining their potential to empower mental health clinicians to better track patients' progress and optimize treatment outcomes with insights from multimodal data. %

To address these gaps and empower mental health clinicians to obtain patient insights from multimodal mental health data, we argue for an alternative approach--designing a \textbf{narrative dashboard}~\cite{bach2023dashboard}, a type of dashboard that presents multimodal data insights in a coherent ``story''. 
As a manifestation, we propose a proof-of-concept system \textbf{\systemName} (Fig.~\ref{fig:teaser}.A)–short for \underline{\textbf{M}}ultimodal data \underline{\textbf{I}}ntegrated \underline{\textbf{N}}arrative \underline{\textbf{D}}ashboard, a computationally curated dashboard that displays multimodal data insights to clinicians through data-driven narratives~\cite{segel2010narrative}. 

To ensure the curation aligned with clinicians' needs, we motivated \systemName by co-designing with five mental health clinicians from varying backgrounds. Through an iterative process, we refined both the design and computation aspects of \systemName and summarized four design goals: intuitive, text-first, adaptive, and insightful (\S\ref{sec:prototyping}).
Then, we implemented \systemName (\S\ref{sec:system}), an interactive interface that displays multimodal insights through data-driven narratives to support various mental healthcare workflows. Organized with a combination of linear (chronological insight organization) and parallel (by Biopsychosocial~\cite{engel1977need, sarafino2014health} model) narrative structures, we used cards as the key means to accommodate multimodal data insights, allowing users to switch between two information levels (Fig.~\ref{fig:system-L1-L2}). To assemble the information used in \systemName, we designed a hybrid computation pipeline (\S\ref{sec:computation}) that pairs the narrative capabilities of large language models (LLMs) with the rigor of rule-based data exploration to curate clinically relevant insights to assist mental health clinicians in better utilizing insights from multimodal patient data. Moreover, we designed the pipeline in a modular way to support future expansion or personalization of the system to more specific workflows.

We evaluated \systemName via a user study (\S\ref{sec:user-study}) with 16 licensed mental health clinicians comparing \systemName with a non-narrative data collection dashboard baseline. Our results reveal that clinicians perceive \systemName can reveal hidden and clinically relevant data insights (\p<.001) and support their decision-making (\p=.004). Additionally, clinicians perceive \systemName as superior to the data collection dashboard baseline in presentation efficiency (\p=.006) and cohesiveness (\p=.004), as well as multimodal data integration (\p<.001).

In summary, our main contributions are as follows:
\begin{itemize}
    \item We describe clinicians' information needs through a co-design study with 5 mental health experts, identifying key concerns when they engage with mental health dashboards.
    \item We propose \systemName, a narrative dashboard curated by a computation pipeline that assists mental health clinicians in obtaining multimodal patient data insights. We open-source our insights generation pipeline and interface design at: \url{https://github.com/sea-lab-space/MIND}
    \item We evaluate \systemName via a user study with 16 licensed mental health clinicians using simulated patient encounter scenarios. We demonstrate how careful computation curation of clinical narratives can empower clinicians in understanding multimodal patient data.
\end{itemize}

Our research demonstrates how multimodal data narratives, paired with carefully designed computation and dashboard design, can empower clinicians to obtain richer information for clinical reasoning. We envision that such narrative approaches can extend beyond mental health and engage other stakeholders (\eg patients, caregivers) to reshape data-driven decision support across healthcare domains (\S\ref{sec:discussion}).

\section{Related Work}
\label{sec:related_work}

We review three lines of work at the intersection of ubiquitous computing, health informatics, and visualization to situate our research. Specifically, we review 1) what data modalities are used in mental healthcare (\S\ref{subsec:related_work:data_modalities}), 2) how multimodal data are currently applied in mental healthcare (\S\ref{subsec:related_work:multimodal-data-use-mh}), and 3) design patterns and practical challenges of clinical dashboards presenting multimodal data (\S\ref{subsec:related_work:dashboards-clinical-practice}).

\subsection{Data Modalities in Mental Healthcare}
\label{subsec:related_work:data_modalities}

The digitization of healthcare over the past decades has gradually transformed healthcare towards more data-driven~\cite{gotz2016datadriven}. Advancements at the intersection of ubiquitous computing and health informatics now enable the collection of multimodal data for mental healthcare. Those multimodal data types could be categorized into clinical data, active sensing data, and passive sensing data ~\cite{adler2025designing}. In this work, we focus on four specific \textbf{data source types} that are either widely accessible in EHRs or increasingly prevalent in mental healthcare contexts: 1) \textit{clinical notes}, 2) \textit{clinical transcripts}, 3) \textit{survey scores}, and 4) \textit{passive sensing data}. 

Clinical notes and transcripts are the two text-based clinical data source types we consider in this study. \textit{Clinical notes} can be retrieved from EHRs~\cite{birkhead2015uses} and are critical to the workflow of mental healthcare practice. These notes include clinician-driven, unstructured source texts that reflect clinicians' assessments and decision-making information about patients. \textit{Clinical transcripts} that contain all conversation details between patients and providers are gradually becoming accessible to more clinicians~\cite{blackley2019speech}. Typically collected through automatic speech recognition, clinical transcripts provide a complementary source to assist clinical documentation.

The other prevailing patient-generated data source types are survey scores and passive sensing data, two time-series, numerical data sources we consider in this study. \textit{Survey scores} are patient self-report mental health measurements collected through active engagement by patients. Those measurements are typically validated methods for evaluating patients' symptoms or functioning, such as PHQ for depression~\cite{kroenke2001phq9, lowe20104item} and GAD for anxiety~\cite{spitzer2006brief,kroenke2007anxiety}. Those scores are collected using methods such as cross-sectional surveys or ecological momentary assessments. In addition to actively collected survey scores, sensors embedded in mobile devices (\eg smartphones, wearables) collect \textit{passive sensing data}, a type of data collected with little human effort (\eg sleep, activity). While those modalities are not direct indicators of mental health conditions, recent research has been exploring data-driven methods to infer mental health symptoms~\cite{dasswain2022semantica}.

Given the complementary nature of the four data types, we investigate how their integration can empower mental health clinicians. In the following sections, we first review the current practices and challenges in leveraging multimodal data, then motivate our approach in integrating multimodal data for mental health clinicians.

\subsection{Leveraging Multimodal Data for Mental Healthcare}
\label{subsec:related_work:multimodal-data-use-mh}

Research has explored multiple ways to leverage multimodal mental healthcare data through both automated and human-centric approaches. For instance, prior work has used transcripts to generate structured clinical notes~\cite{schloss2020automated, lin2018reimagininga, han2024ascleaia}. To reduce the complexity of these notes, studies have also proposed automated summarization methods~\cite{mishra2014text, afantenos2005summarization} and visual analytics systems~\cite{sultanum2018morea, sultanum2019doccurate, sultanum2023chartwalk}, enabling clinicians to review and interpret information at scale. One prevailing paradigm for utilizing patient-generated data is to develop models that ``detect/predict''~\cite{adler2024detectiona} mental health measures such as mood~\cite{morshed2019prediction, li2020extraction}, stress~\cite{yu2023semisupervised, egilmez2017ustress, hovsepian2015cstressa}, and depression~\cite{xu2019leveraginga, xu2021leveraginga, wang2018trackinga}. However, recent evidence suggests that such models often fail to generalize across datasets~\cite{xu2023globema, adler2022machineb, meegahapola2023generalization} and are typically centered on enabling low-burden methods for continuous symptom monitoring outside the clinic~\cite{adler2022call, insel2017digital}. Few have discussed how patient-generated multimodal data can be used inside of clinics and linked to clinical notes and transcripts, to provide clinicians with insights that support in-session treatment decision-making beyond prediction~\cite{adler2024detectiona}.

In this work, we make an initial attempt to thread together all four data source types as complementary information to support mental health clinicians. Inspired by recent research~\cite{englhardt2024classificationc, heydari2025anatomy, khasentino2025personal}, we propose a LLM-powered pipeline that translates multimodal data into data insights. We further embed the data insights in a narrative dashboard to facilitate clinicians' consumption and reasoning with multimodal text insights.

\subsection{Designing Dashboards for Clinical Practice}
\label{subsec:related_work:dashboards-clinical-practice}

Effectively communicating the rich information embedded in health data is an active research topic~\cite{shneiderman2013improving}. Dashboards, ``\textit{a visual display of data used to monitor conditions and/or facilitate understanding}''~\cite{wexler2017big, sarikaya2019whata}, emerged as a promising solution because of its potential to allow clinicians to grasp complex information ``\textit{at a glance}''~\cite{few2006information, bach2023dashboard}. Specifically, we focus on clinical dashboards~\cite{dowding2015dashboards}, a subset of healthcare dashboards focusing on informing clinicians with relevant data with the aim to improve patient care quality (\eg~\cite{powsner1994graphical}).

Researchers have designed clinical dashboards for multiple data types (see reviews~\cite{dowding2015dashboards, helminski2024development, zhuang2022framework, shenvi2023visualization}), including clinical notes~\cite{sultanum2018morea, sultanum2019doccurate, sultanum2023chartwalk, vanderlinden2023medicospace}, time-series/event sequence data~\cite{powsner1994graphical, shahar2000modelbased, shahar2006distributed, sharmin2015visualizationb, wang2008aligninga, kandel2024pdinsighter, elm2019feasibilityb, sadhu2023designingb}, domain-specific collection of multiple data types~\cite{kim2017prescribinga, evans2024using, jiang2024healthprism, wu2025cardioai, yao2025more}. While these dashboards effectively support targeted workflows, many remain single-modal (\eg focusing only on notes) or adopt a data collection dashboard design pattern--dashboards that present high volumes of information with minimal curation of what and how the data is presented~\cite{bach2023dashboard}. For example, CardioAI~\cite{wu2025cardioai} integrates passive sensing with voice assistant data to summarize cardiotoxicity risk, yet presents the two modalities in separate modules. Such layouts expose all relevant information but risk increasing clinicians' cognitive burden, particularly in time-sensitive clinical environments where mental capacity is already stretched~\cite{zeldes2011information, ye2021impact}.

To address this limitation, we explore an alternative paradigm--designing narrative dashboards~\cite{bach2023dashboard}. Drawing from narrative visualization research~\cite{segel2010narrative}, we integrate a ``storyline'' that threads together multimodal data. Perhaps the most relevant research to our work is from \citet{zhang2013fivea}, which proposed using the Five Ws (who, when, what, where, and why) from journalistic reporting to visualize patient information. We extend this perspective by applying linear and parallel narrative techniques to communicate multimodal mental health insights for clinicians. Further, building on calls to center text in clinical visualization~\cite{sultanum2018morea}, we leverage the expressive potential of LLM-generated insights (as evidenced in \S\ref{subsec:related_work:multimodal-data-use-mh}) to communicate multimodal data. We expect the combination of multimodal data, algorithmic insight curation, and narrative presentation to provide mental health clinicians with a complementary view they can leverage in their daily workflow~\cite{adler2024detectiona}.

\section{\systemName Design}
\label{sec:prototyping}

\begin{table*}[t]
\centering
\small
\caption{Experts participated in the co-design process.}
\Description{
Domain experts who participated in the co-design process. The group included five professionals (one clinical psychologist, three psychiatrists, and one research psychologist) with diverse expertise across digital mental health, substance abuse, ADHD, and cannabis use, as well as strong research backgrounds. Their years of experience ranged from 6 to over 20 years, ensuring that feedback reflected both clinical practice and research perspectives.
}
\begin{tabular}{cllll}
\toprule
\textbf{E\#} & \textbf{Gender} & \textbf{Profession} & \textbf{Area of Expertise} & \textbf{Years of Experience} \\ \midrule
E1 & Female & Clinical Psychologist & Technology for mental health, Researcher & 16-20 years\\
E2 & Male & Psychiatrist & Digital mental health & 7-10 years\\
E3 & Male & Psychiatrist & Mental health, Substance abuse & 7-10 years\\
E4 & Male & Research Psychologist & Digital mental health, Depression, Researcher & 20+ years\\
E5 & Male & Psychiatrist & ADHD, Cannabis use & 11-15 years\\
\bottomrule
\end{tabular}
\label{tab:co-design-experts}
\Description{}
\end{table*}

While prior works discussed clinical dashboard designs in multiple clinical settings (\S\ref{subsec:related_work:dashboards-clinical-practice}), less is known about mental health clinicians' needs in making sense of multimodal data across clinical and patient-generated data~\cite{li2025vital} and how to design systems to support this task. To execute our research in a principled way, we adopted a process following the design study methodology~\cite{sedlmair2012design}. We chose this methodology because it's proven for developing visualization systems that address real-world tasks, and has been widely applied in HCI and health-related systems design (\eg~\cite{kandel2024pdinsighter, kwon2019retainvis}).

To characterize the needs of mental health clinicians, we conducted a co-design study with five mental health experts.  
We focus on the four clinically meaningful data types identified in \S\ref{subsec:related_work:data_modalities}: 1) clinical notes \inlinefig{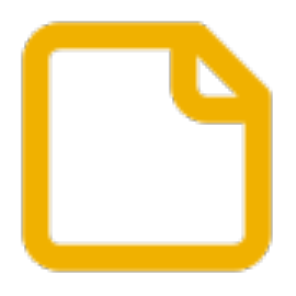}, 2) clinical transcripts \inlinefig{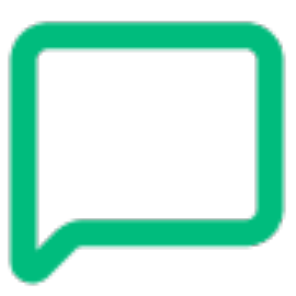}, 3) survey scores \inlinefig{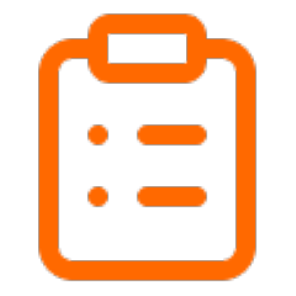}, and 4) passive sensing data \inlinefig{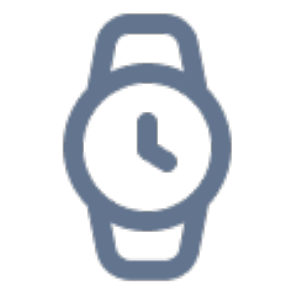}.
In the following, we first describe our co-design process (\S\ref{subsec:codesign-process}). Then, we report insights from expert feedback (\S\ref{subsec:codesign-insights}) and outline four design goals (\S\ref{subsec:codesign-implications}) that guided the development of \systemName.

\subsection{Co-designing with Experts}
\label{subsec:codesign-process}

We engaged five domain experts (E1-E5) in psychiatry and psychology (Tab.~\ref{tab:co-design-experts}) to bridge the mental health domain with the technical areas and collaboratively design \systemName~\cite{sedlmair2012design}. We recruited experts through a combination of convenience and purposive sampling~\cite{etikan2016comparison} to ensure they represented perspectives from diverse mental health clinical settings (\eg academic medicine, digital health innovation, private practice, and global mental health). We conducted multiple co-design sessions with experts spanning over two months to ensure \systemName targets real-world clinical needs. We envisioned \systemName to be mainly used for patient case review at multiple stages in the clinical workflow, often in situations where decision time is limited, \eg preparation (patient recap prior to the clinical encounter) and consultation (decision-making during clinical sessions).

\begin{figure*}[t]
    \centering
    \includegraphics[width=.9\linewidth]{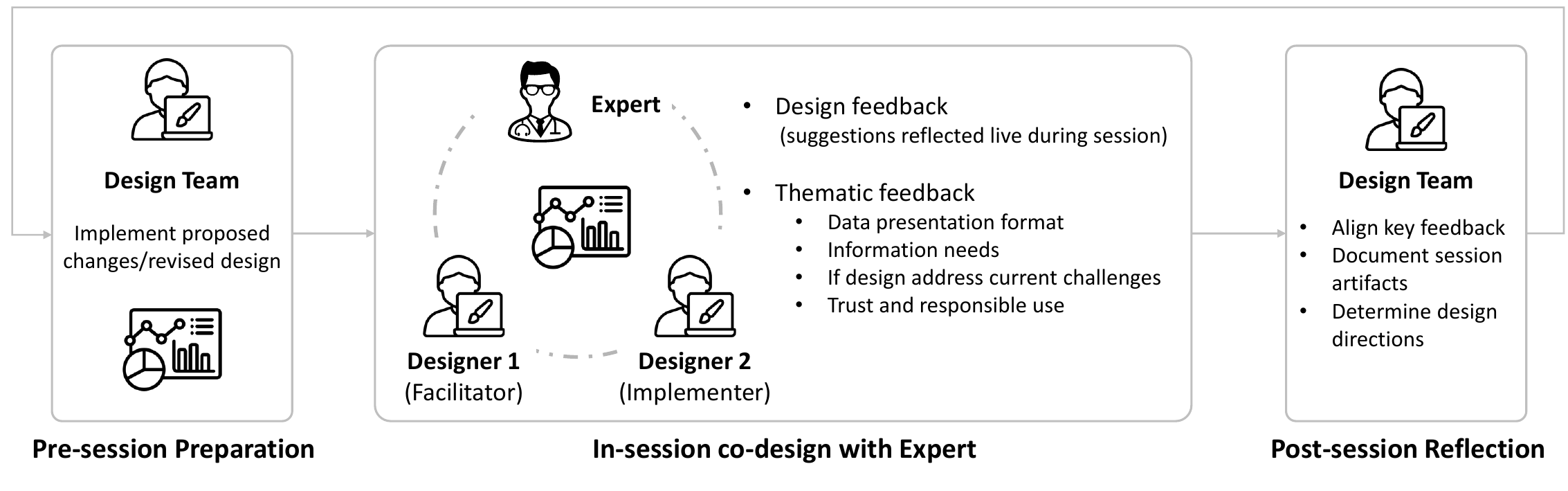}
    \caption{Illustration of the procedure of the co-design process.}
    \Description{The co-design process contains three steps: pre-session preparation, in-session co-design with expert, and post session reflection. In pre-session preparation, the design team implement proposed changes and revised design. In the in session co-design process, designer 1, the facilitator, and designer 2, the implementer, interacts with the expert on the dashboard design. They seek feed back regarding design, and thematic. Thematic feedback include data presentation format, information needs, if the design addresses current challenges, and trust and responsible use of AI and dashboard. In the post-session reflection, the design team align key feedback, document key session artifacts, and determine next step design directions.}
    \label{fig:co-design-process-flow-chart}
\end{figure*}

Before the co-design process, we reviewed prior literature and created an initial dashboard. Then, in co-design sessions (approximately 1 hour each), we followed an unstructured, exploratory interview process (Fig.~\ref{fig:co-design-process-flow-chart}) and invited experts to propose and refine prototypes together alongside designers on a version of the prototype (updated based on prior co-design sessions). We scheduled each expert individually for focused design sessions, and two authors (the designers) of the paper participated in each session. During each session, one author (\ie Facilitator) facilitates the discussion, and the other takes notes and implements real-time updates to prototypes (\ie Implementer). Besides detailed suggestions, we elicit feedback on broader topics such as their preferred data presentation format (\ie text, charts) for different tasks (\eg overview, fact-checking), information needs (\ie what needs to be shown on the dashboard), constraints in current dashboards/information systems, and how to promote trust for clinicians using an AI-powered dashboard. We instantiate larger revisions between sessions to reflect the broader thematic suggestions.

Throughout the co-design, we created seven early iterations of \systemName by rapid prototyping in Figma\footnote{\url{https://www.figma.com/}} and developing interactive interfaces through programming. Fig.~\ref{fig:iteration-progress} illustrates four design checkpoints that marked the progression of design (see Appendix~\ref{appendix:artifacts-codesign} for prototype screenshots). \systemName evolved from a chart-heavy dashboard design towards a more text-centered, narrative dashboard design. The earliest prototypes (Fig.~\ref{fig:iteration-progress}.C1, C2) relied on stacked or tiled text-chart pairings that clinicians found visually overwhelming and lacked sufficient logic. Specifically, C1's heavy emphasis on charts remains hard for mental health clinicians to interpret. While C2 improved upon C1 by integrating more text, it remains scattered and without a clear focal point. Building on C1 and C2, C3 (Fig.~\ref{fig:iteration-progress}.C3) introduced a narrative structure that grouped text descriptions of data insights. Welcoming this change, clinicians continued to request clearer navigation and categorization given the large number of insights. The final checkpoint, C4 (Fig.~\ref{fig:iteration-progress}.C4), improved upon C3 by providing components that enable easier user control in navigating text descriptions of data insights. Moreover, after we finalized the implementation, we additionally ran two pilot sessions with E1 and E3 with the full-functioning \systemName dashboard and finalized the design.

\subsection{Design Process Findings}
\label{subsec:codesign-insights}

Synthesizing feedback across the co-design process, we distill takeaways regarding mental health experts' needs in viewing multimodal data. We refer to design checkpoints (C\#) to provide context to these findings. %

\subsubsection{Clinicians' workflows differ across specialties, but common ground exists at the conceptual level.}
Clinicians proposed multiple workflows working with multimodal data. 
We found common approaches that clinicians use to make sense of multimodal data. In earlier iterations (C1, C2, C3), experts often proposed adding time-related information about the timeframe used to derive insights, indicating that they make sense of a patient's progress by looking at a sequence of events. Another common approach is through categorization, first proposed as a feature by E2. Early prototypes (C2, C3) took E2's suggestion by organizing information by thematic topics (\eg sleep, activities, emotions). While experts appreciate the feature, they did not reach consensus on which topics should be included because of differences in clinical practice. 
To provide a common ground for all clinicians, E5 suggested adopting recognized conceptual frameworks such as the Biopsychosocial model~\cite{engel1977need, sarafino2014health}, an established framework in the mental health domain, as a familiar baseline for categorization. Although not all experts used the model in practice, they found it easy to understand and appropriate for navigating insights(C4). Such feedback suggested that we should organize information through workflow-agnostic structures or established mental health frameworks.

\subsubsection{Clinicians prefer text in most cases but value visual aids for navigation and information seeking.}

Early prototypes were primarily designed around charts (C1), but several experts (E1, E3, E5) emphasized that text is the preferred format to convey insights.
While charts were considered a powerful way of conveying information, they were less familiar to mental health clinicians and were often treated as secondary information aids. E4 suggested presenting textual insight descriptions as structured summaries (\eg bullet points), and using charts only as a form of providing supplementary information on demand to avoid cluttering the interface. Regarding the textual insight description, E1 stressed the importance of using clear, simple language rather than confusing technical terminology. E2 and E5 also recommended using icons and color-coding as information cues. Such feedback suggested that we should make text a central, navigable part of the dashboard, while allowing for detail-on-demand via charts.

\begin{figure*}[t]
    \centering
    \Description{
    Design evolution of multimodal data dashboards. We iteratively developed four candidate designs for presenting clinical insights:
    C1. Linked Text/Chart Stacks: textual summaries and visualizations displayed in separate vertical stacks, connected by highlighting.
    C2. Tiled Text-Chart Pairs: insights presented as side-by-side text–chart pairs to support immediate comparison.
    C3. Three-stage Narrative: structured around three layers—overview, detailed clinical insights, and patient communication—to align with clinical reasoning.
    C4. Linear + Parallel Narrative with Navigation: the final design, combining linear storytelling (medical history, session recap, patient data insights, and summary) with parallel navigation and filtering, enabling both holistic and detailed review.
    This progression illustrates how co-design feedback shaped the system from simple pairings of text and charts into a narrative-driven dashboard with flexible navigation and information depth.
    }
    \twiceinclude[width=.96\linewidth]{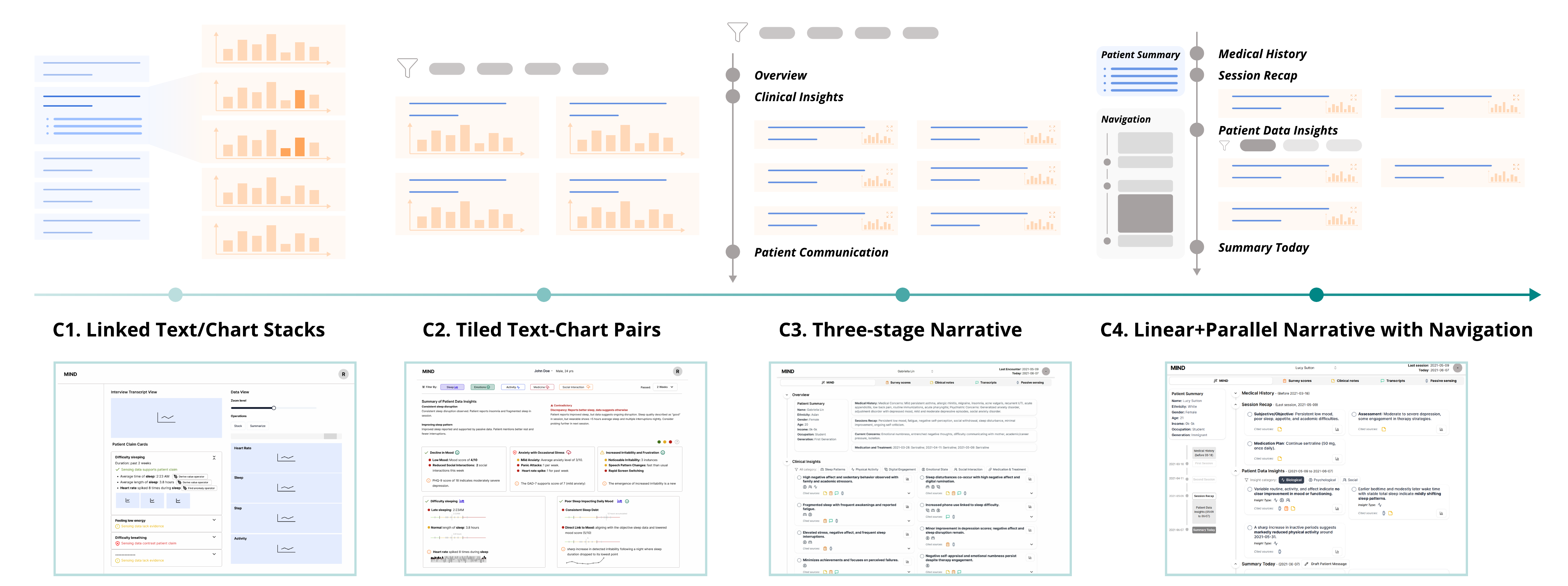}
    \Description{Domain experts who participated in the co-design process. The group included five professionals (one clinical psychologist, three psychiatrists, and one research psychologist) with diverse expertise across digital mental health, substance abuse, ADHD, and cannabis use, as well as strong research backgrounds. Their years of experience ranged from 6 to over 20 years, ensuring that feedback reflected both clinical practice and research perspectives.}
    \caption{Four checkpoint designs (C1-C4) showing the co-design progression incorporating experts' ideas and feedback. The top row illustrates the design pattern, and the bottom row presents the \textcolor{captionGreen}{prototype artifacts} that instantiate each pattern. Information layout becomes more structured, while \textcolor{captionOrange}{charts} become more closely integrated with \textcolor{captionBlue}{text} content with each iteration. We present expanded views of the prototype artifacts for C1 (Fig.~\ref{fig:appendix:system-initial}), C2 (Fig.~\ref{fig:appendix:system-pre-nar}), and C3 (Fig.~\ref{fig:appendix:system-after-nar}) in the Appendix.}
    \label{fig:iteration-progress}
\end{figure*}

\subsubsection{Clinicians value simple interfaces given time constraints in clinical practice.}
Clinical workflows are highly time-sensitive, particularly during pre-session preparation. E5 noted that clinicians often only have about five minutes to review and recap a patient's information, and E2 emphasized that back-to-back sessions leave little time for a deep review. Such reports echoed prior findings (\eg~\cite{sultanum2018morea}) that demonstrate a typically 2-10 minutes preparation stage note reviewing time and extended it into the mental healthcare settings. In light of the time constraints, prototypes (C1, C2) were frequently critiqued as overwhelming, with experts describing the excessive amount of text and charts as contributing to cognitive overload. Particularly, interfaces that displayed too much raw data made it difficult for clinicians to retrieve the most relevant insight, especially under a short timeframe. Such feedback suggested that we should display a selected subset of information at first glance to avoid cognitive overload.

\subsubsection{Clinicians acknowledge AI could be a useful tool, with caveats in their role and reliability.} 
During the co-design process, clinicians expressed nuanced expectations for how AI should be integrated into the dashboard. While they agreed that AI could be useful for summarizing subtle data patterns that might otherwise be overlooked (E2), they cautioned against diagnostic or prescriptive uses. For example, E3 emphasized that AI-generated suggestions should be phrased in a non-diagnostic manner, while E4 supported AI recommending potential activities based on patient symptoms, but only with clinicians' approval. Experts also raised concerns about AI-generated content impacting or even replacing their professional judgment, noting that auto-generated insights risked preventing clinicians' cognitive engagement. To leverage the benefit and maintain clinicians' agency, they proposed several design ideas to keep the transparency and traceability of the design--generated insights should link back to raw data sources so clinicians could verify and challenge the system when necessary. Overall, clinicians envisioned AI as an assistive tool that surfaces insights and suggestions while they preserve the agency in diagnosis and shared decision-making with patients. Such feedback suggested that we should generate clinically relevant insights in a reliable, traceable way to foster trust in the dashboard.

\subsection{Design Implications}
\label{subsec:codesign-implications}

By summarizing takeaways through co-designing with clinicians, we derived four design guidelines:

\begin{enumerate}
    \item[\textbf{DG1}] \textbf{Intuitive: Design intuitive and at-a-glance clear information presentation.} \systemName should allow clinicians to easily locate and retrieve relevant clinical information to accommodate the time-sensitive clinical workflow.

    \item[\textbf{DG2}] \textbf{Text-first: Apply text-first design to enable efficient information uptake.} \systemName should prioritize concise and simple textual summaries over jargonistic content or complex visualizations, and apply visual cues to improve the navigation of the text.

    \item[\textbf{DG3}]
    \textbf{Adaptive: Structure information to adapt to different information needs.} \systemName should provide detailed information on demand to supplement the text-based data insights for fact-checking and trust-building.

    \item[\textbf{DG4}] %
    \textbf{Insightful: Generate reliable, clinically relevant data insights.} The pipeline that generates clinician-facing content should ensure the generation of data insights relevant to clinicians' workflow, while keeping the process reliable and traceable. Besides, constraints should be applied to the pipeline to ensure insights are concise, clinically appropriate, and free of ambiguous or overly technical phrasing. %

\end{enumerate}

\section{\systemName Interface}
\label{sec:system}

To validate the design goals summarized through the co-design process, we created \systemName, a multimodal data-integrated narrative dashboard. Following the visual information-seeking mantra~\cite{shneiderman1996eyes},
\textit{``overview first, zoom and filter, details on demand''}, we designed \systemName with two information levels. Accordingly, L1 (\S\ref{subsec:narrative-structure}, Fig.~\ref{fig:system-design}) narrates multimodal data insights (DG4)--data patterns interpreted to describe a patient's condition--with established narrative structures (DG1) and text descriptions (DG2). Meanwhile, L2 (\S\ref{subsec:L2-level}, Fig.~\ref{fig:system-L1-L2}) displays data facts, or ``\textit{patterns, relationships, or anomalies extracted from data under analysis}''~\cite{chen2009effective} that support the data insights for trust building and data exploration (DG3).

\subsection{L1: Narrating Multimodal Data Insights}
\label{subsec:narrative-structure}

\systemName's L1 interface consists of two components: Patient Background for basic information and navigation, and Narrative Overview to show the data insights. Specifically, we employed two narrative structures (\ie linear and parallel storytelling) and sought to strike a balance between information communication and discovery (\ie balancing user- and author-driven data stories)~\cite{segel2010narrative, lan2021understanding, yang2022design, lee2015more,bach2023dashboard}.

\begin{figure*}[tb]
    \centering
    \twiceinclude[width=.89\linewidth]{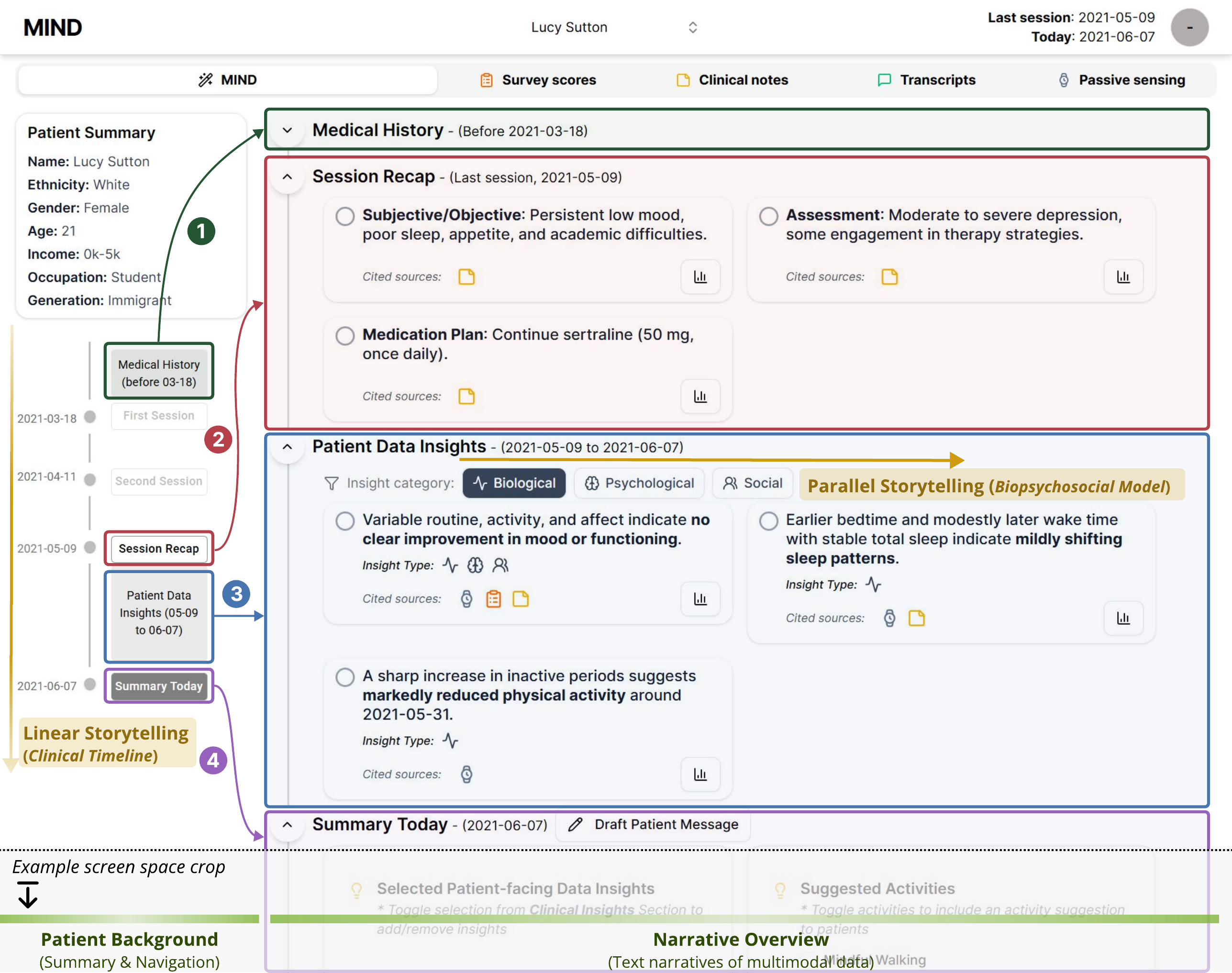}
    \Description{The L1 overview interface of MIND. The system communicates multimodal data insights by combining linear storytelling and parallel storytelling techniques. On the outer level, MIND follows a four-part linear structure—Medical History (collapsed), Session Recap, Patient Data Insights, and Summary Today—organized along a clinical timeline to mirror patient review workflows. Within the Patient Data Insights section, MIND employs parallel narrative threads based on the biopsychosocial model, allowing clinicians to facet insights across biological, psychological, and social domains. Supporting components include a Patient Background panel for demographic context and navigation, and a Details-on-Demand layer enabling clinicians to expand summarized insights into source-linked data. Together, these features support efficient case review while maintaining access to detailed clinical evidence.}
    \caption{The L1 overview interface of \systemName. \systemName communicates multimodal data insights through a combination of linear and parallel storytelling techniques. On the outer level, \systemName narrates insights using a linear four-part structure: \textcolor{teaserGreen}{Medical History (collapsed) \ding{182}}, \textcolor{teaserRed}{Session Recap \ding{183}}, \textcolor{teaserBlue}{Patient Data Insights \ding{184}}, and \textcolor{teaserPurple}{Summary Today \ding{185}}, with overflowing content accessible through scrolling\inlinefig{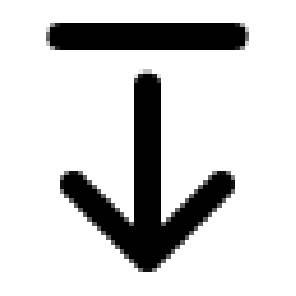}. Within the Patient Data Insights section, \systemName further applies three parallel narrative threads based on the Biopsychosocial model for clinician-guided information faceting.}
    \label{fig:system-design}
\end{figure*}

The first information point clinicians see is the \textbf{Patient Background} situated at the left of the interface, which presents 1) a summary of the patient's demographic data and 2) a clinical timeline that illustrates the series of mental health treatment sessions conducted with the clinician. From the clinical timeline, clinicians can navigate across the dashboard--each button on the timeline corresponds to a section in the Narrative Overview component. We use the components in this column as the primary hub for ``\textit{messaging}''~\cite{segel2010narrative}, communicating both the summary clinical information and the overall organization of the dashboard.

The main content of \systemName lies in \textbf{Narrative Overview}, situated at the right of the interface. To ensure the dashboard is intuitive and efficient (DG1, DG2), we took an ``\textit{author-driven}'' approach~\cite{segel2010narrative} by choosing chronology to bind the story that enforces a specific order in presenting the data insights. Specifically, laid out following the \textit{linear narrative} (\ie in the exact chronological order of how events unfold), clinicians can make sense of the patient's data by reading sequentially through the four sections (S1-S4): Medical History (Fig.~\ref{fig:system-design}.\ding{182}, the narrative's ``setting''), Session Recap (Fig.~\ref{fig:system-design}.\ding{183}, the narrative's ``setting''), Patient Data Insights (Fig.~\ref{fig:system-design}.\ding{184}, the narrative's ``rising-climax''), and Summary Today (Fig.~\ref{fig:system-design}.\ding{185}, the narrative's ``resolution''). Such a sequence also closely aligns the key narrative stages identified by \citet{yang2022design}, who applied Freytag's pyramid narrative structure~\cite{freytag1895technique} to data stories. While research in narrative visualization has shown the potential of using anachronies (\ie present events in an order that deviates from the exact chronology) to improve user engagement in reading data stories~\cite{lan2021understanding}, experts in our co-design session championed the timeline design, which we argue is more common~\cite{hullman2013deeper} and anticipated in clinical practice.

Regarding interaction, we expand the Session Recap section by default and collapse other sections to focus on clinical notes insights--a common starting point for patient review for mental health clinicians. Clinicians can click the button next to the section titles or the button on the timeline to expand\inlinefig{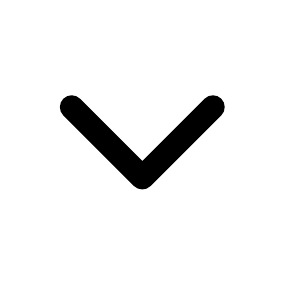} or collapse\inlinefig{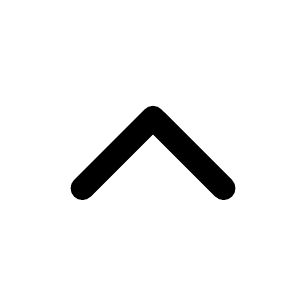} each section. Following the typical design of a narrative dashboard, we keep every insight on the same page and display overflow content in the Narrative Overview component with scrolling~\cite{bach2023dashboard}.
In the following, we first introduce the visual primitive that contains the data insights. Then, we break down the four sections in the Narrative Overview and explain the rationale behind the design.

\subsubsection{Visual Primitive: Insight Card}

\begin{figure}[tbp]
    \centering
    \twiceinclude[width=.8\linewidth]{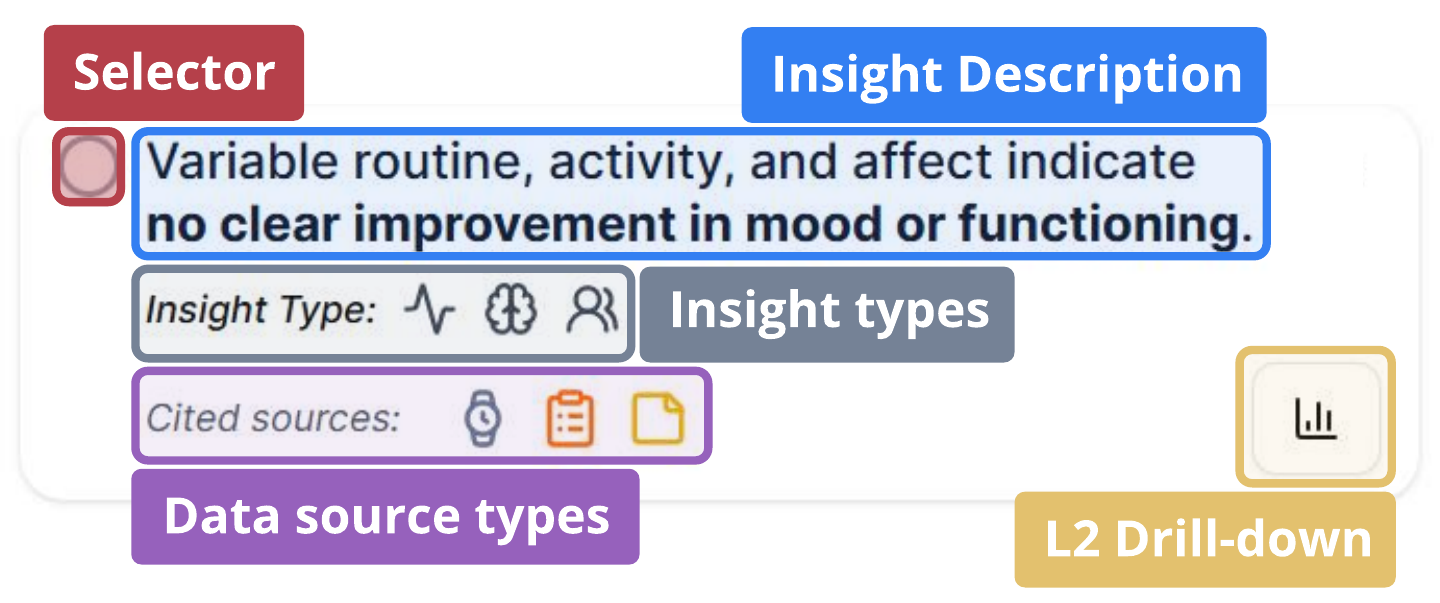}
    \caption{Insight card as visual primitive. Three sections: insight description, insight types, and data source types, present the multimodal insights. The selector and drill-down buttons links the card with other parts of the \systemName.}
    \Description{
    Anatomy of a patient data insight card in MIND. Each card contains: a selector (for inclusion in patient-facing summaries), a concise insight description generated from multimodal data, and insight type icons (e.g., biological, psychological, social). Cards also cite their data source types (e.g., passive sensing, clinical notes, transcripts), supporting traceability to underlying evidence. A Level-2 drill-down control enables expansion into detailed visualizations and raw data for clinician verification.
    }
    \label{fig:insight-card}
\end{figure}

We present multimodal data insights within \textbf{Insight Cards} (Fig.~\ref{fig:insight-card}). Specifically, we lay out key insight information: textual \textit{insight description}, the \textit{insight types} (biological~\inlinefig{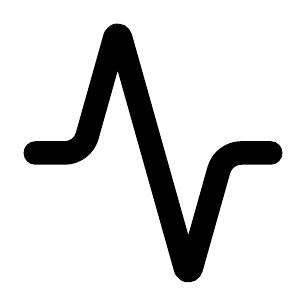}, psychological~\inlinefig{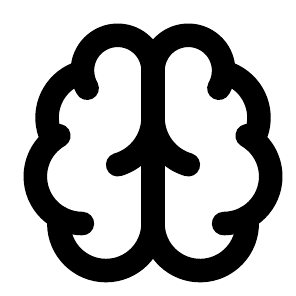}, and social~\inlinefig{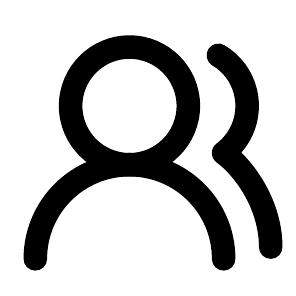}), and the four \textit{data source types} (\ie \inlinefig{passivesensing}~\inlinefig{activesensing}~\inlinefig{clinicalnote}~\inlinefig{clinicaltranscript}) used to generate the insight in the center.
Both insight types and data source types are coded in icons, and the data source types are further color-coded to highlight the multimodality of the data insight. Two clickable buttons, the \textit{selector} \inlinefig{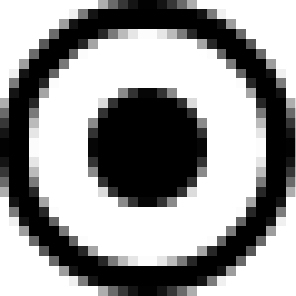} and the \textit{drill-down} \inlinefig{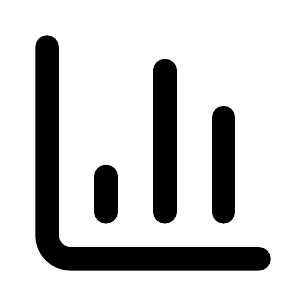}, link the data insight with the Summary Today section in L1 and the drill-down panel in L2, respectively.

\subsubsection{S1: Medical History}
The Medical History section contains summative descriptions of a patient's medical history beyond the ongoing mental health treatment sessions.
We included the patient's past physical and psychological history with time markers indicating the onset of those symptoms (Fig.~\ref{fig:system-design}.\ding{182}).

\subsubsection{S2: Session Recap} 
Session Recap contains a three-piece summary from a past session note (Fig.~\ref{fig:system-design}.\ding{183}). Drawing on the structure of a widely applied note format, SOAP (Subjective, Objective, Assessment, Plan)~\cite{sarafino2014health}, we present summaries in three insight cards with prefixes: Subjective/Objective, Assessment, and (medication) Plan. We merge the S and O components because both capture observations of the patient's state. We keep the Plan section strictly limited to medication-related matters. Clinicians can add the insight to the S4: Patient Communication Section or also drill down to see the raw notes through the drill-down button (see \S\ref{subsec:L2-level}). 

\subsubsection{S3: Patient Data Insights}
Building on the session recap, Patient Data Insights (Fig.~\ref{fig:system-design}.\ding{184}) shows ``\textit{what happened since the last visit}''~\cite{sultanum2018morea}. Within the section, we employ the Biopsychosocial model~\cite{engel1977need, sarafino2014health}--a framework that breaks down mental health causes into biological, psychological, and sociological factors--to structure data insights through \textit{parallel narrative} (\ie multiple separated stories linked to a common theme). Specifically, we selected the Biopsychosocial model because it is 1) suggested by experts in our co-design study, 2) widely known by mental health clinicians~\cite{mahapatra2024biopsychosocial}, and 3) carves an appropriate amount of ``sub-stories'' (three) for clinicians to explore. Each ``story'' is about one of the three facets in the Biopsychosocial model~\cite{sarafino2014health}. Clinicians are given the affordance to select which ``storyline'' they want to see with a filter function with three options: Biological~\inlinefig{bio}, Psychological~\inlinefig{psyc}, and Social~\inlinefig{social}.

Within each parallel story, we laid out insight cards in a grid-like, two-column setting. We argue that current capabilities of computation can not decide ``the most important'' insight for clinicians. As a heuristic, we ordered data insights that used more data sources before single-modal data insights, and left the final prioritization and interpretation to clinicians (\ie designing ``\textit{user-driven}'' data stories~\cite{segel2010narrative}). Regarding data insight description, we employed a two-part template: 1) factual descriptions of the patient's condition (\eg ``earlier bedtime'') plus 2) how it implies the patient's condition (\eg ``shifting sleeping patterns'', bolded to foster efficient reading). Similar to the Session Recap section, clinicians can add the insight to the S4: Patient Communication Section or drill down to see the raw data facts that support the insight.

\subsubsection{S4: Summary Today}
The Summary Today section (Fig.~\ref{fig:system-design}.\ding{185}) offers an open-ended conclusion to the Session Recap (S2) and Patient Data Insight (S3) by showing all selected data insights to provoke post-session patient communication. Based on the context, clinicians can select several suggested activities and use the \textit{Draft Patient Message} function to create a text message that integrates patient data insights and suggested activities for communication.

\begin{figure*}[tbp]
    \centering
    \twiceinclude[width=0.9\linewidth]{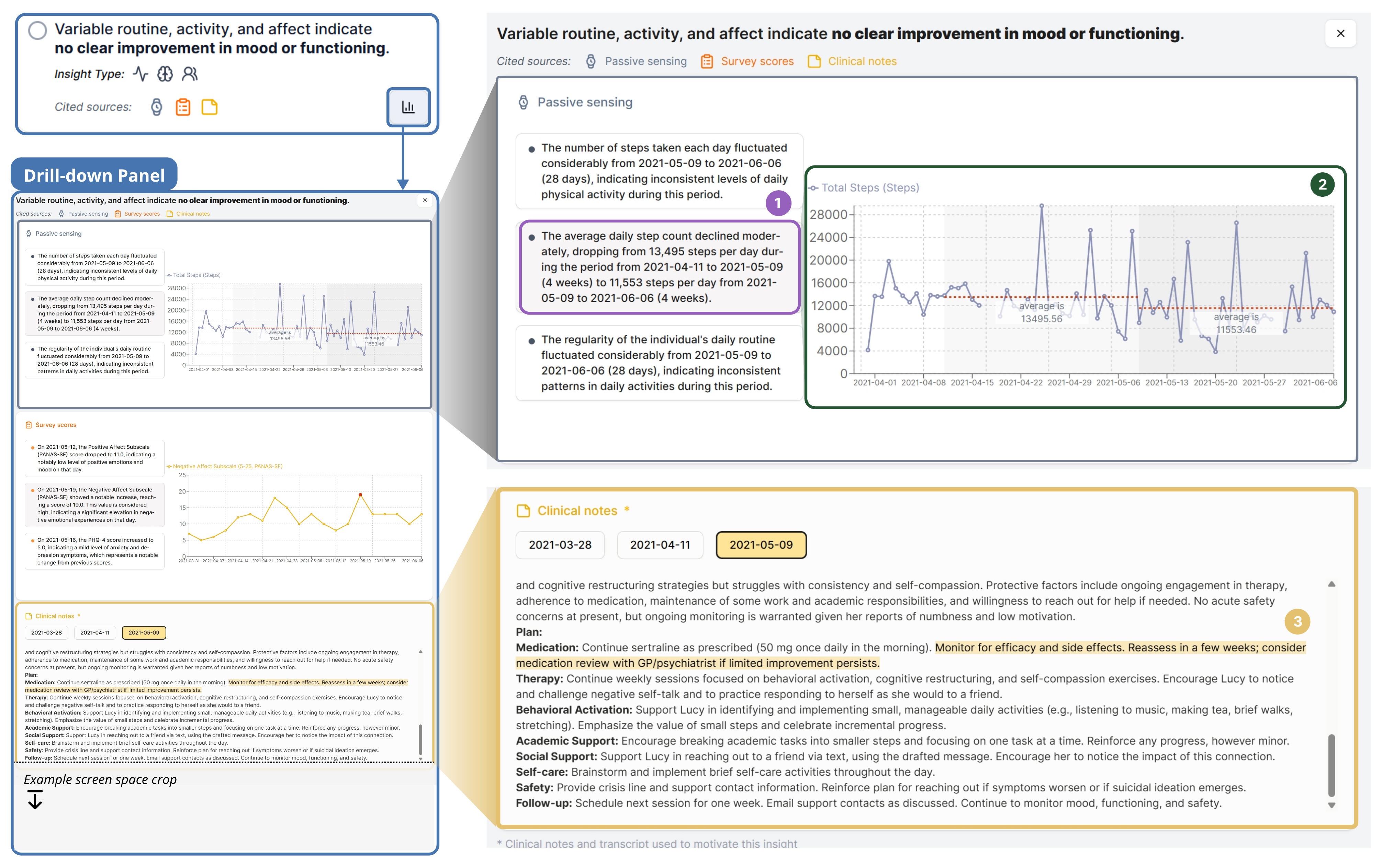}
    \caption{Drill-down panel for a patient data insight. Users can open the panel by clicking the drill-down button in L1. Each section provides detailed information for clinicians to explore underlying data, and each \textcolor{teaserPurple}{\ding{182} summarized bullet point} corresponds with a \textcolor{teaserGreen}{\ding{183} chart} supporting the data fact. \textcolor{captionYellow}{\ding{184} Text highlight} shows the text piece most relevant to the data insight.}
    \Description{
    Drill-down panel for patient data insights. Clicking the chart button expands an insight card into a drill-down view that organizes supporting evidence across multiple data modalities: passive sensing, survey scores, clinical notes, and transcripts. Summarized bullet points highlight key observations from passive sensing, each linked to a corresponding graph visualization, while clinical notes and transcripts provide contextual narrative evidence. This layered design enables clinicians to move seamlessly from high-level summaries to granular, source-linked details, supporting transparency and trust in the generated insights.
    }
    \label{fig:system-L1-L2}
\end{figure*}

\subsection{L2: Data Facts on Demand}
\label{subsec:L2-level}

The L2 drill-down panel included detailed displays of data facts that support the data insights in L1. Clinicians can click on the drill-down button from the insight cards (Fig.~\ref{fig:system-L1-L2}). Inside the drill-down panel, we display the data insight description on top, followed by the data source types. Subsequently, we categorize data facts by their data source type: passive sensing, active sensing (survey scores), clinical notes, and clinical transcripts. 
Additionally, we put numerical data first to give clinicians an immediate, quantitative view, then textual data source types as supporting evidence.

To illustrate the data facts from passive and active sensing data sources, we pair each data fact (Fig.~\ref{fig:system-L1-L2}.\ding{182}) with a chart (Fig.~\ref{fig:system-L1-L2}.\ding{183}). We created visual elements on the charts (\eg average value, trend line) to clearly demonstrate the key data attribute described in the data fact. Moreover, we chose simple chart types to minimize the learning curve from clinicians while following best practices from the visualization community (\eg using a linear y-axis scale; using the most appropriate chart type: line charts to display data trend/bar charts to show differences and extremes) to accurately reflect the underlying data and mitigate potential data-driven biases.

To show insights inspired by clinical notes and transcripts, we highlight the text that contributes to each insight to enable efficient skimming (Fig.~\ref{fig:system-L1-L2}.\ding{184}). We also show the full history of clinical data to support further backtracking, while highlighting the current date for quick navigation to the most recent date. Moreover, the drill-down panel provides a critical guardrail, enabling clinicians to rigorously ascertain the correctness of the presented insights.

\subsection{Implementation}
We implement \systemName using a typical web stack, using React\footnote{\url{https://react.dev/}} as the UI framework. To render the charts used in \systemName, we used Recharts\footnote{\url{https://recharts.org/}} for its built-in interactive data visualizations. For the content, the system consumed a JSON data structure passed from the computation pipeline (\S\ref{sec:computation}).

\section{\systemName Computation Pipeline}
\label{sec:computation}

To curate the content for \systemName, we implemented a proof-of-concept pipeline to transform multimodal data into reliable, clinically relevant data insights \textbf{(DG4)}. In this section, we first describe how we formulate the computation tasks and utilize the multimodal data (\S\ref{sub:computation-data-usage}). Then, we provide descriptions of the design of each module in our pipeline (\S\ref{sub:computation-summarizer}, \S\ref{sub:computation-insights}). Last, we introduce the technical implementation details of the pipeline (\S\ref{sub:computation-implementation}).

\begin{figure*}[bp]
    \centering
    \includegraphics[width=0.75\linewidth]{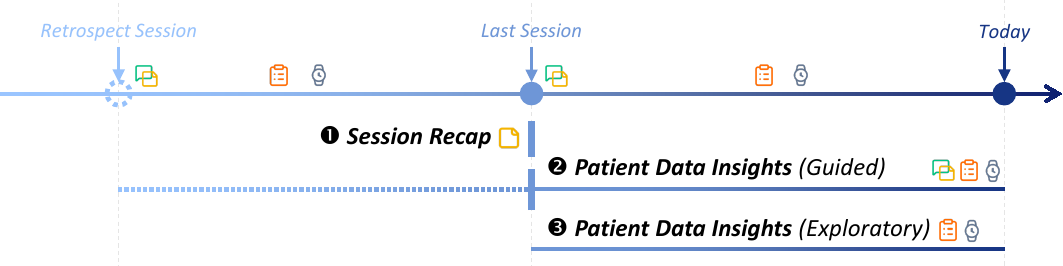}
    \caption{Data modalities used for generating each insight thread. \ding{182} Session Recap uses clinical note \inlinefig{clinicalnote} from the last session. \ding{183} The Structured Patient Data Insights thread combines clinical data (last session) \inlinefig{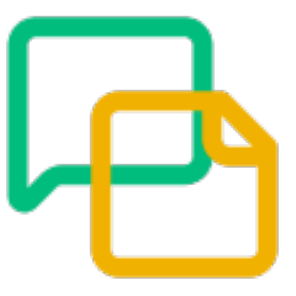} with new patient-generated sensing data \inlinefig{activesensing}\inlinefig{passivesensing}, using historical data for context. \ding{184} The Exploratory Patient Data Insights thread analyzes only new patient-generated sensing data \inlinefig{activesensing}\inlinefig{passivesensing} since the last session.}
    \Description{
    Data modalities underlying different narrative threads in MIND. The Session Recap is generated from clinical data recorded in the last session. The Structured Patient Data Insights thread combines last session’s clinical data with new sensing data (surveys, passive sensing) to provide contextualized interpretation. The Exploratory Patient Data Insights thread focuses exclusively on new sensing data since the last session, supporting the detection of emerging patterns independent of prior clinical notes. This layered use of modalities allows clinicians to flexibly review insights anchored in both historical context and recent patient experiences.
    }
    \label{fig:pipeline-timeline}
\end{figure*}

\subsection{Insight Discovery Tasks, Data Usage, and Method}
\label{sub:computation-data-usage}

First, we define the types of insights needed and the time points or intervals associated with their computation. For each upcoming session, we consider generating information for two sections in the L1 Narrative Overview component: 1) session recap (\ie what was discussed last session, Fig.~\ref{fig:system-design}.\ding{183}), and 2) patient data insights (\ie how the patient's condition has evolved between the sessions, Fig.~\ref{fig:system-design}.\ding{183}). We define three time points: 1) \textit{today} (\ie the upcoming session date), 2) \textit{last session} (the previous session date), and \textit{the retrospective sessions} (the session before the last session). Fig.~\ref{fig:pipeline-timeline} shows how each data modality contributes to session recap and patient data insights. 

First, we discover session recap insights only using the last session's clinical notes \inlinefig{clinicalnote} (Fig.~\ref{fig:pipeline-timeline}.\ding{182}) because they are clinician-driven narratives that reflect what they deemed important about the patient. Meanwhile, we decided not to include sensing data \inlinefig{activesensing}\inlinefig{passivesensing} for the session recap because they are too remote for the current date.

Second, we discover patient data insights using all four data source types \inlinefig{clinicalnote}\inlinefig{clinicaltranscript}\inlinefig{activesensing}\inlinefig{passivesensing}. For clinical notes and transcripts, we use them to inform data insight discovery (\eg analyze sleep data if mentioned in clinical notes).
We refer to this process as \textit{guided discovery} (Fig.~\ref{fig:pipeline-timeline}.\ding{183}). In parallel, relying solely on sensing data \inlinefig{activesensing}\inlinefig{passivesensing}, we designed an \textit{exploratory discovery} (Fig.~\ref{fig:pipeline-timeline}.\ding{184}) process to capture ``anomalies''~\cite{li2025vital} that may be neglected by structured discovery.

Within our pipeline, all modules described are driven by a large language model (LLM) agent unless explicitly stated. Specifically, we followed best practices in building LLM applications (\eg using zero-shot Chain-of-Thought prompting~\cite{kojima2022largea}, prompt-chaining techniques~\cite{wu2022ai}) to optimize the performance of our pipeline.

\begin{figure*}[b]
    \centering
    \twiceinclude[width=.97\linewidth]{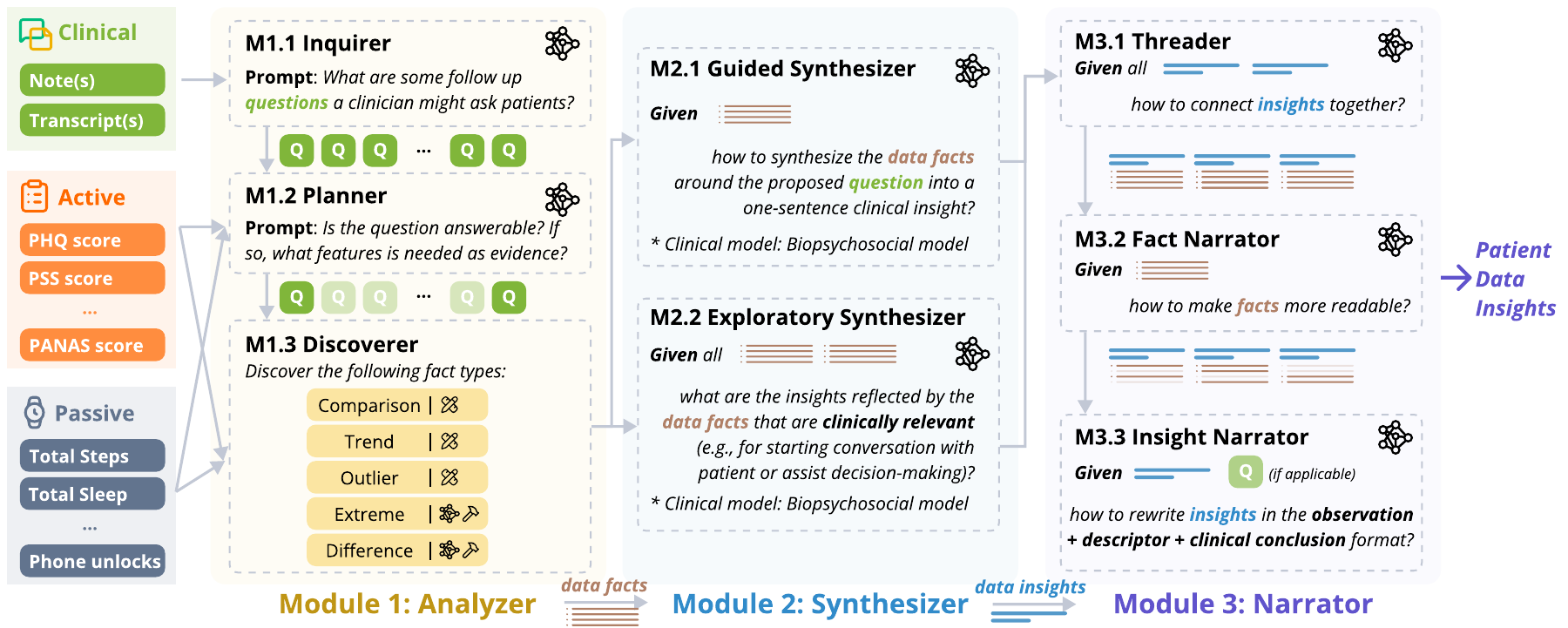}
    \caption{The patient data insight discovery pipeline of \systemName contains three modules: \textcolor{discovererColor}{Analyzer (\textbf{M1})}, \textcolor{synthesizerColor}{Synthesizer (\textbf{M2})}, and \textcolor{narratorColor}{Narrator (\textbf{M3})}. We display the computation method used in each module/submodule using icons, with \inlinefig{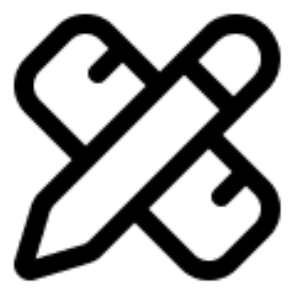} indicating rule-based approaches, \inlinefig{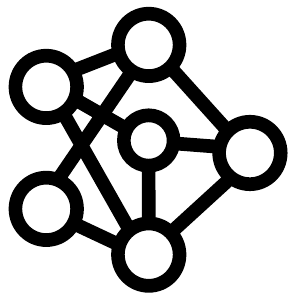} indicating LLM-based approaches, while \inlinefig{llm}\inlinefig{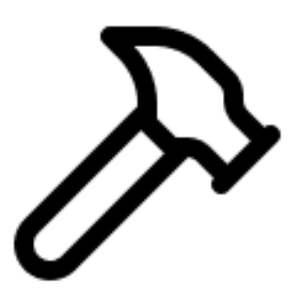} indicating tool-using-LLM-based approaches.}
    \Description{
    The patient data insight discovery pipeline of MIND consists of three sequential modules: Analyzer (M1), Synthesizer (M2), and Narrator (M3). In the Analyzer module, the system begins with the Inquirer, which generates potential follow-up questions that a clinician might ask based on the available patient data. These questions are then processed by the Planner, which determines whether a question is answerable and identifies the features required as evidence. Finally, the Discoverer identifies and extracts fact types such as comparisons, trends, outliers, extremes, and differences across clinical, active sensing, and passive sensing data. These outputs form the foundational data facts. In the Synthesizer module, the system builds on these facts to produce clinically relevant insights. The Guided Synthesizer organizes facts around specific questions and applies the biopsychosocial model to contextualize them, while the Exploratory Synthesizer examines all data facts to generate new observations that may be useful for starting clinical conversations or supporting decision-making. In the Narrator module, the system translates these synthesized insights into structured narratives for clinician use. The Threader connects related insights together into coherent groups, the Fact Narrator reformulates raw data facts into more readable text, and the Insight Narrator rewrites insights into a structured format that combines the observation, a descriptive element, and a clinical conclusion. Together, these three modules create a pipeline that moves from raw multimodal patient data through fact extraction and synthesis to the generation of concise, clinically interpretable narratives.
    }
    \label{fig:computation}
\end{figure*}

\subsection{Insight Discovery for Session Recap}
\label{sub:computation-summarizer}

We generate the Session Recap section with \textbf{Summarizer}. Summarizer takes in raw clinical notes and transforms them into a modified SOAP structure specified in \S\ref{subsec:narrative-structure}. For each insight item, we constrain the LLM to generate insights of no more than 12 words to keep them succinct. Moreover, we instruct the LLM to provide evidence on which pieces of clinical notes are used to create the insight summary to support the text highlight feature (\S\ref{subsec:narrative-structure}).

\subsection{Insight Discovery for Patient Data Insight}
\label{sub:computation-insights}

We employed a hybrid approach that combines rule-based and LLM-based methods. Through three modules \textbf{Analyzer} (\S\ref{sub:sub:computation-discoverer}), \textbf{Synthesizer} (\S\ref{sub:sub:computation-synthesizer}), and \textbf{Narrator} (\S\ref{sub:sub:computation-narrator}), the pipeline takes the four data source types \inlinefig{clinicalnote}\inlinefig{clinicaltranscript}\inlinefig{activesensing}\inlinefig{passivesensing} as input and outputs data insight-fact pairs. The system was architected using a pipes-and-filter software architecture (Fig.~\ref{fig:computation}) to minimize dependencies across modules. We also indicate whether each component module is rule-based~\inlinefig{rulebase} or LLM-based~\inlinefig{llm} with icons.

\subsubsection{Insight Generation Framework}

\renewcommand{\arraystretch}{1.2}
\begin{table*}[htbp]
\small
  \centering
  \caption{Template-based data fact generation rules used by the \textit{\textbf{Analyzer}} module. We specify each fact type with its temporal feature (single time point/time duration, $t$/$\Delta t$), value ($v$/$\Delta v$), and a value-derived attribute ($a$). We include the entity related to the data fact ($\{e\}$, \eg ``total sleep'') in the description syntax to provide context for a data fact.}
  \Description{
  Template-based data fact generation rules are employed by the Analyzer module. Each fact type is specified by a temporal dimension (either a single time point or a duration), a value representation, and a value-derived attribute. These components are then mapped into a natural language template to produce a fact description that clinicians can interpret directly. For example, the Outlier type identifies anomalies such as spikes or dips at a particular time, while the Trend type captures directional or cyclical changes across a period. Comparison facts contrast average values between two intervals, whereas Difference facts describe changes in magnitude between two points in time. Finally, Extreme facts highlight the maximum or minimum observed values. In all cases, the entity being measured (for instance, total sleep, physical activity, or survey score) is inserted into the template syntax to ground the fact in a specific clinical or behavioral variable. This template-based approach ensures consistency in how raw multimodal data are translated into structured data facts, forming the foundation for higher-level synthesis and narrative generation.
  }
    \begin{tabularx}{\linewidth}{lcccX}
    \toprule
    \textbf{Fact Type} & \textbf{Time ($t$/$\Delta t$)} & \textbf{Value ($v$/$\Delta v$)} & \textbf{Attribute ($a$)} & \textbf{Description Syntax} \\
    \midrule
    Outlier & $\Delta t$ & $v$ & \makecell[c]{spike/dip} & An anomaly was detected for $\{e\}$ on $\{t\}$, which $\{a\}$ to $\{v\}$. \\
    Trend & $\Delta t$ & - & \makecell[c]{\xspace \\  rise/fall/stable/\\cyclic/variable/\\no (trend) \\ \xspace} & The $\{e\}$ showed $\{a\}$ trend from $\{\Delta t_1\}$ to $\{\Delta t_2\}$. \\
    Comparison & $\Delta t_1$, $\Delta t_2$ & $\bar{v}_{\Delta t_1}$, $\bar{v}_{\Delta t_2}$ & \makecell[c]{increase/decrease/\\remained stable} & The average $\{e\}$$\{a\}$ from $\{\bar{v}_{\Delta t_1}\}$ ($\Delta t_1$) to $\{\bar{v}_{\Delta t_2}\}$ ($\Delta t_2$). \\
    \midrule
    Difference & $t_1$, $t_2$ & $v_{t_1}$, $v_{t_2}$ & \makecell[c]{more/less} & The $\{e\}$ was $\{v_{t_1}\}$ on $\{t_1\}$ and became $\{a\}$ at $\{v_{t_2}\}$ on $\{t_2\}$. \\
    Extreme & $t$ & $v$ & \makecell[c]{max/min} & The $\{e\}$ reached its $\{a\}$ value of $\{v\}$ on $\{t\}$. \\
    \bottomrule
    \end{tabularx}
  \label{tab:mind-fact-type}
\end{table*}
\renewcommand{\arraystretch}{1.0}

Taking inspiration from automatic visualization research, we employ the data fact taxonomy~\cite{chen2009effective} for a principled approach in managing the data fact discovery process. Specifically, the data fact taxonomy provides formal definitions to categorize data insights (\eg, value, distribution, difference) in visual analytics systems. Recent research has successfully applied this taxonomy in several computational approaches to extract insights from multimodal datasets (\eg~\cite{zou2025gistvis, chen2024chart2vec, shi2021calliope, wang2020datashot}).

While the original fact taxonomy included 12 data fact types, we consider a subset of five data fact types most relevant to our task: \textit{comparison}, \textit{trend}, \textit{outlier}, \textit{extreme}, and \textit{difference}. Specifically, the first three data fact types measure data attributes across one or multiple time periods ($\Delta t$ or $\Delta t_1, \Delta t_2$), whereas the latter two measure data attributes at single time points (\textit{extreme}: $t$; \textit{difference}: $t_1, t_2$). Given the data fact types and their temporal attributes, we define a \textit{fact} as an intermediate layer between raw data and data facts. We formulate a \textit{fact} as a 4-tuple:
\begin{equation}
\mathTextItalics{fact}\, \aptLtoX[graphic=no,type=html]{\coloneqq}{\coloneq} \left \{ \mathTextItalics{type, Time, Value, attribute} \right \} \nonumber
\end{equation}
where \textit{type} is one of the five fact types we selected. Meanwhile, \textit{Time} and \textit{Value} are tuples that define the temporal feature of data facts. Combining \textit{type} and \textit{value}, \textit{attribute} is a derived feature describing the characteristics of the data fact. Tab.~\ref{tab:mind-fact-type} documents the Time, Value, and Attribute of the five fact types we consider in this study.

\subsubsection{Analyzer}
\label{sub:sub:computation-discoverer}

Analyzer (Fig.~\ref{fig:computation}.~M1), transforms raw multimodal data into \textit{facts}. To derive data facts grounded with clinical data, we first process clinical notes and transcripts through two agents: Inquirer (Fig.~\ref{fig:computation}.~M1.1) and Planner (Fig.~\ref{fig:computation}.~M1.2). Analyzer reflects on clinical data, especially the forward-facing plans, to generate a set of questions aligned with mental health clinicians' interests. Then, we pass all passive and active sensing data features to the Planner to determine 1) if the proposed question is answerable, and 2) if answerable, what features the Analyzer should investigate. Those filtered questions are then sent to the  Discoverer (Fig.~\ref{fig:computation}.~M1.3). To ensure that the extracted data facts are grounded in raw data, we designed Discoverer with two complementary mechanisms: 1) rule-based agents for extracting comparison, trend, and outlier data facts, and 2) LLM-based exploratory agents with tool-assisted verification for extreme and difference data facts.

First, taking as input the \textit{guided} plan from the Analyzer and Planner, we designed a series of rule-based agents to discover outlier, trend, and comparison data facts. Because the discovery goal is to reveal any significant changes between sessions, the time intervals are defined as the period from the last session to today. We applied statistical and heuristic methods to discover data facts from raw data. For comparison, we apply the Mann-Whitney U test to uncover differences in medians between two time periods split by the last session (see Fig.~\ref{sub:computation-data-usage}). For trend detection, we use the Mann-Kendall test to identify monotonic increases or decreases, complemented by autocorrelation analysis for cyclic detection and the coefficient of variation method for stability assessment. For outlier detection, we apply seasonal-trend decomposition via LOESS (locally estimated scatterplot smoothing) and identify anomalies in the residual component using the Median Absolute Deviation rule, classifying them as spikes or dips. 

Second, complementing the guided discovery, we used LLMs for \textit{exploratory} discovery on extreme and difference data facts. Grounded by LLMs' potential in understanding sensing data~\cite{englhardt2024classificationc, heydari2025anatomy, khasentino2025personal}, we provided raw \texttt{csv} to the LLM-based Discoverer (Fig.~\ref{fig:computation} M1.3) agents to capture data facts. Additionally, to mitigate potential hallucination~\cite{huang2025survey}, we guardrail the agent with a data verification tool: whenever a data retrieval task is required, the LLM invokes a predefined script to access raw values directly from the dataset, minimizing the instances where the LLMs might ``invent'' raw data.

Crucially, both the \textit{guided} and \textit{exploratory} discovery processes are agnostic to the underlying active and passive sensing measurements. Designers can tailor the discovery process with a specific set of measurements best suited to specific mental health cases or clinicians. Moreover, both discovery processes are also robust to missing data. Missing data are either handled directly by methods that tolerate incomplete observations (\eg Mann-Whitney U and Mann-Kendall tests) or linearly interpolated over time when required (\eg autocorrelation analysis and anomaly detection). We preserve the raw data (\ie include data missing) in the output to the \systemName interface to indicate missing data to users.

After the hybrid LLM/rule-based approach transformed numerical data into fact specifications, we applied a rule-based natural language generation method to convert these data facts into data fact descriptions. Extracted fact specifications fill in the blanks we designed for each data fact type (Tab.~\ref{tab:mind-fact-type}, Description Syntax column). Rather than relying on code representations (\ie JSON specification), we prepared data facts as natural language descriptions to fully leverage the LLM's language understanding capabilities for subsequent insight synthesis. We also specifically opted for this controlled process to ensure that the description of data facts is accurate and not altered during LLM's natural language generation.

\subsubsection{Synthesizer}
\label{sub:sub:computation-synthesizer}

Taking the guided and exploratory data facts, \textbf{Synthesizer} keeps the two types of generation and combines data facts.
Guided Synthesizer (Fig.~\ref{fig:computation}.~M2) surface data insights from data facts that potentially answer each computable question proposed in \textbf{Analyzer}. If significant data facts exist, guided synthesizers only take significant results as input for data insight synthesis. If no significant data facts exist, we instruct the Guided Synthesizer agent to consume all data facts and generate an insight using up to six data facts. Meanwhile, the Exploratory Synthesizer (Fig.~\ref{fig:computation}.~M2.2) surface anomalies observed through one large corpus of data facts. Specifically, we instruct the Exploratory Synthesizer to output at least 15 ``clinically relevant'' insights (\ie assist clinical decision-making or good material to raise a patient discussion in a session).

To ensure that each data insight could be traced back to data facts, we instructed both synthesizer modules to output the data fact evidence alongside the synthesized data insight, creating data insight-facts pairs. Moreover, we steered both synthesizer agents to consider the Biopsychosocial model during synthesis and labeled each insight with one or multiple categories in the Biopsychosocial model. To keep final data insights succinct and readable, we constrain both synthesizer agents to generate insights of fewer than 15 words using plain language.

\subsubsection{Narrator}
\label{sub:sub:computation-narrator}

The \textbf{Narrator} module (Fig.~\ref{fig:computation}.~M3) combines the data insight-facts pairs from the \textbf{Synthesizer} module into clinician-aligned content. First, the Threader (Fig.~\ref{fig:computation}.~M3.1) combines the insights from the two synthesizers into a coherent narrative. We keep all insights from Guided Synthesizer (M2.1) because they are clinically grounded and use a combination of sensing and clinical data. We add four to six insights from the Exploratory Synthesizer (M2.2) to balance between guided and exploratory insights.

Taking the full patient data insights as input, the Fact Narrator (Fig.~\ref{fig:computation}.~M3.2) and Insight Narrator (Fig.~\ref{fig:computation}.~M3.3) optimize content quality. Specifically, the fact narrator processes the template-generated data fact description from Analyzer and paraphrases it to make it more readable. While this approach did make the data fact description longer, experts considered the additional details appropriate for describing the data fact. Subsequently, the insight narrator further rewrites the data insights to align with the clinician's information needs using the template introduced in \S\ref{subsec:narrative-structure}. 

It is worth noting that despite our LLM generation guardrails, errors might still occur from Synthesizer (\eg synthesizing clinically unrelated data facts) and Narrator (\eg semantic shifts when rewriting data facts/insights). To mitigate this risk, we designed the L2 drill-down panel to be immediately accessible (\ie the drill-down button) to the data insights, prompting clinicians to verify insights that seemed nonsensical to them.

\subsection{Implementation and Internal Validation}
\label{sub:computation-implementation}
We implemented our computation pipeline in Python. The pipeline communicates the generated content with the frontend via a JSON file, analogous to a live backend server. For LLMs, we used OpenAI's \textsc{GPT-4.1} API\footnote{\url{https://openai.com/api/}}--a non-reasoning model optimized for building agentic applications. To enable tool integration, we used the OpenAI Agents SDK\footnote{\url{https://openai.github.io/openai-agents-python/}}, allowing the LLM to call functions and retrieve raw data.
We used packages such as \texttt{statsmodels}, \texttt{scipy}, and \texttt{pymannkendall} for the statistical tests in the rule-based agents.

Following the design study methodology~\cite{sedlmair2012design}, we conducted multiple rounds of ``inward-facing'' validations of our computational pipeline. Specifically, two mental health experts (E1 and E2) independently reviewed multiple batches of LLM-generated summaries and data insights to ensure the appropriateness of wording and the factuality of the generated insights. Specifically, experts evaluated whether the wording avoided diagnostic claims, whether the language was easy to read, and whether each insight related to the raw data. For instance, reviewing the sentence ``\textit{Sleep disturbances co-occur with high negative affect and digital rumination}'', E2 requested the insights to be more descriptive. We thus modified our computation pipeline to favor wording such as ``\textit{Variable bedtime and awakening episodes with stable total sleep and wake time indicate fluctuating sleep quality}.'' Experts also flagged cases where insights were not clearly grounded in raw data, and we subsequently applied more stringent feature selection requirements. Overall, their feedback helped refine our templates, constraints, and evidence-linking mechanisms. We terminated the validation process after both health experts agreed upon the generated content. %

\section{User Evaluation}
\label{sec:user-study}

To evaluate if and how well \systemName supports mental health clinicians in obtaining insights from multimodal data, we conducted a mixed-method within-subject study with 16 licensed mental health clinicians in a simulated patient review task\footnote{Approved by the last author's institute's IRB.}. We compared \systemName with a baseline system \baselineName that employs the data collection dashboard design that displays the same multimodal data as \systemName. We want to investigate:
\begin{itemize}
    \item \textbf{Quality}: How do clinicians evaluate the quality (\eg clinical relevance) of the generated data insights in \systemName?
    \item \textbf{Perception}: How do users perceive the narrative dashboard design of \systemName, and how does it impact system usability and perceived workload?
\end{itemize}

\begin{table*}[t]
\small
  \centering
      
    \caption{Demographics of user evaluation participants. We use SV, PS, CN, and CT to denote survey scores~\inlinefig{activesensing}, passive sensing data~\inlinefig{passivesensing}, clinical notes~\inlinefig{clinicalnote}, and clinical transcripts~\inlinefig{clinicaltranscript}, respectively. We measure participants' attitude towards AI on a scale of 1 to 10, with a higher score indicating a more positive attitude. }
    \Description{
    A table titled "Demographics of user evaluation participants" details the profile of 16 participants (P1-P16). The table has 6 columns: participant ID, gender, practice focus, years of experience, reported data experience, and AI attitude. Most of the participants are female, with practice focus covering Psychiatry, Counseling, Social Work, Clinical Psychology, and Behavior Analysis. Participants have varying years of experience and display different experiences with data. 13 out of 16 have experience in survey data (SV), 10 with passive sensing data (PS), 16 with clinical notes (CN), and 9 with clinical transcripts (CT). Overall, clinicians have an AI attitude score of 8.06 out of 10.
    }

    \begin{tabular}{lllllllll}
    \toprule
    \multirow{2}[0]{*}{\textbf{P\#}} & \multirow{2}[0]{*}{\textbf{Gender}} & \multirow{2}[0]{*}{\textbf{Practice Focus}} & \multirow{2}[0]{*}{\textbf{Years of Experience}} & \multicolumn{4}{l}{\textbf{Reported Data Experience}} & \textbf{AI Attitude}~\cite{grassini2023development} \\
          &       &       &       & SV~\inlinefig{activesensing} & PS~\inlinefig{passivesensing}    & CN~\inlinefig{clinicalnote}   & CT~\inlinefig{clinicaltranscript} & (1-10, negative-positive)\\

    \midrule      
    P1    & Female & Psychiatry & 7-10 years & \textbullet     & \textbullet     & \textbullet     &  & \progressplot[6]{6.5}{1}{10}\\
    P2    & Female & Psychiatry & 7-10 years & \textbullet     & \textbullet     & \textbullet     &  & \progressplot[6]{7}{1}{10}\\
    P3    & Female & Psychiatry & 7-10 years & \textbullet     &       & \textbullet     & \textbullet & \progressplot[6]{8.75}{1}{10}\\
    P4    & Male  & Counseling, Social Work & 10+ years & \textbullet     & \textbullet     & \textbullet     & \textbullet & \progressplot[6]{7.5}{1}{10} \\
    P5    & Female & Counseling & 1-3 years & \textbullet     & \textbullet     & \textbullet     & \textbullet & \progressplot[6]{7.75}{1}{10}\\
    P6    & Female & Behavior Analyst & 10+ years & \textbullet     &       & \textbullet     & \textbullet & \progressplot[6]{10}{1}{10}\\
    P7    & Male  & Clinical Psychology & 10+ years & \textbullet     &       & \textbullet     &  & \progressplot[6]{9.5}{1}{10}\\
    P8    & Male  & Counseling & 4-6 years &       &       & \textbullet     &  & \progressplot[6]{5.75}{1}{10}\\
    P9    & Female & Counseling, Social Work & 1-3 years &       &       & \textbullet     &  & \progressplot[6]{10}{1}{10}\\
    P10   & Female & Psychiatry & 1-3 years & \textbullet     & \textbullet     & \textbullet     & \textbullet & \progressplot[6]{9}{1}{10}\\
    P11   & Female & Counseling & 4-6 years & \textbullet     & \textbullet     & \textbullet     & \textbullet & \progressplot[6]{8.25}{1}{10}\\
    P12   & Female & Counseling & 1-3 years &       &       & \textbullet     &  & \progressplot[6]{9}{1}{10}\\
    P13   & Male  & Psychiatry & 10+ years & \textbullet     & \textbullet     & \textbullet     &  & \progressplot[6]{9.25}{1}{10}\\
    P14   & Female & Clinical Psychology & 7-10 years & \textbullet     & \textbullet     & \textbullet     & \textbullet & \progressplot[6]{8.5}{1}{10}\\
    P15   & Female & Clinical Psychology & 4-6 years & \textbullet     & \textbullet     & \textbullet     & \textbullet & \progressplot[6]{6}{1}{10}\\
    P16   & Female & Psychiatry, Social Work & 10+ years & \textbullet     & \textbullet     & \textbullet     & \textbullet & \progressplot[6]{6.25}{1}{10}\\
    \midrule
    \multicolumn{4}{l}{\textbf{Summary}} & 13 & 10 & 16 & 9 & \progressplot[6]{8.06}{1}{10}\\
    \bottomrule
    \end{tabular}%
    \label{tab:user-testing-participant}

\end{table*}

\subsection{Participants}
\label{sub:user-study:participants}

We recruited 16 participants (12 self-identified as female, 4 as male) who did not participate in our co-design process through a combination of convenience, purposive, and snowball sampling~\cite{etikan2016comparison, goodman1961snowball} (Tab.~\ref{tab:user-testing-participant}). All participants are licensed mental health practitioners with at least 1 year of experience, among whom 5 have over 10 years of experience in mental healthcare. Participants cover various practices, among which 6 reported expertise in psychiatry, 6 in counseling, 3 in clinical psychology, and 3 in social work. Most participants (14 of 16) reported using an EHR system regularly as part of their clinical practice. All participants reported working with clinical notes while using different data modality combinations in their daily practice. Specifically, 7 have prior experience in all four data categories, while 3 have only worked with clinical notes. Using the AI Attitude Scale (AIAS-4)~\cite{grassini2023development}, participants reported overall positive attitude in AI (\textit{M}=8.06, \textit{SD}=1.42).

\subsection{Materials}
\label{sub:user-study:materials}

To simulate real-world scenarios, we prepared three simulated users: Lucy, Gabriella, and Alison, synthesizing data streams from multiple real-world datasets in collaboration with our experts (E1-E5). We utilized the GLOBEM dataset~\cite{xu2022globema}, an open-source longitudinal dataset that includes both passive and active sensing data from real-world users between 2018 and 2021. Using Beck's Depression Inventory (BDI-II)~\cite{beck1996beck} in the dataset, we screened out users with moderate or high depression symptoms ($\text{BDI-II} \geq 19$) as archetype simulated patients. Then, by manually going through their other EMA survey scores (\eg PHQ-4~\cite{lowe20104item}, PANAS~\cite{watson1988development}), we manually sampled three typical instances as simulated patient archetypes.

Meanwhile, since GLOBEM was designed primarily for ML model training and evaluation, we picked a subset of easy-to-understand features for materials preparation and the later evaluation studies. Specifically, we retained all six regularly collected active sensing data features (see Appendix \S\ref{appendix:ps-survey}) and 11 passive sensing data features spanning four categories (Location, Screen, Sleep, and Step, see full list in Appendix \S\ref{appendix:ps-data}). 

Subsequently, sourcing publicly available mental health session tutorial videos, we picked three instances to present to our experts as baselines that we pair with the three patient archetypes. Then, for each patient archetype, we worked with our experts in \S\ref{sec:prototyping} and paired each archetype with a tutorial session. Subsequently, we created two subsequent mock interview sessions with three clinical notes (see pairing in Appendix \S\ref{appendix:synthetic-data-generation}). Overall, while our setup does not cover all ecological setups (\ie other note formats or interview strategies exist), we aim to provide a homogeneous baseline across all three simulated patients to minimize the noise caused by different study materials and focus on evaluating the efficacy of the narrative design of \systemName.

\begin{figure*}
    \centering
    \includegraphics[width=.95\linewidth]{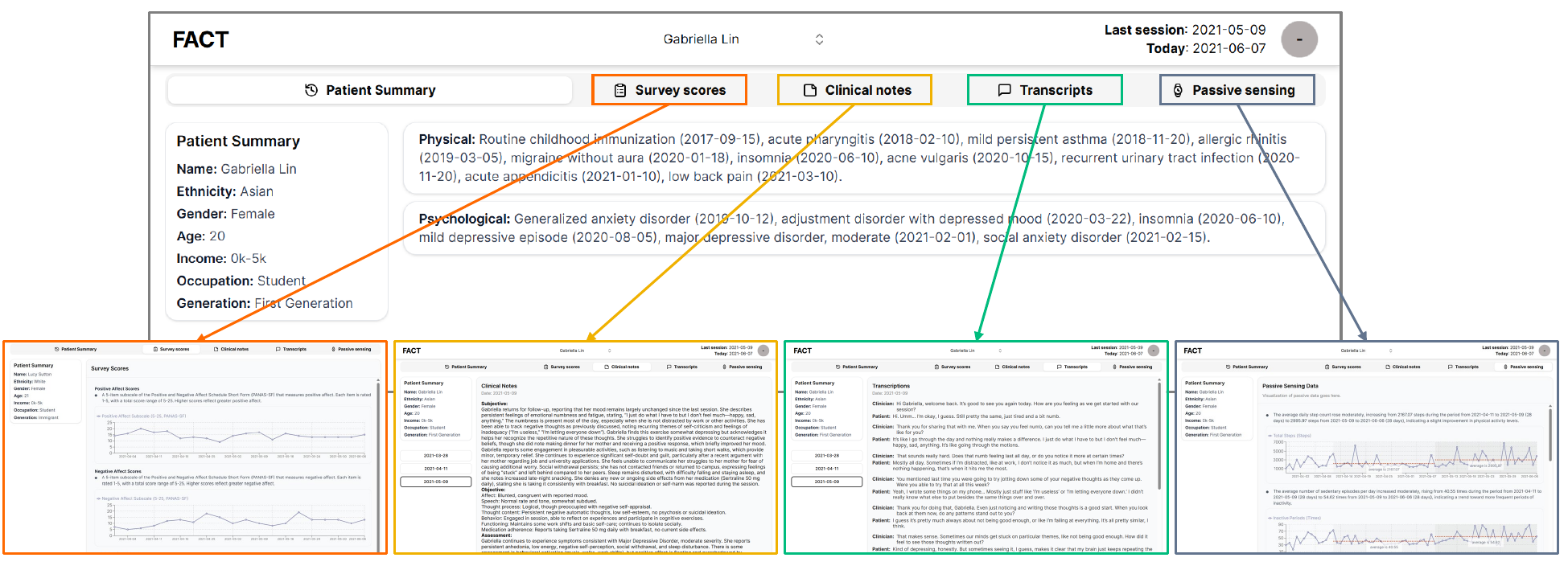}
    \caption{The baseline interface \baselineName. \baselineName displays the same multimodal data as \systemName, but in four separate views (tabs): Survey scores, Clinical notes, Transcripts, and Passive sensing.}
    \Description{The baseline interface, FACT, which was used for comparison with MIND in the evaluation study. FACT presents the same multimodal patient data—survey scores, clinical notes, transcripts, and passive sensing—but organizes them into four separate tabbed views. Each tab isolates a single data modality, requiring clinicians to switch between views to access different types of information. For example, survey results are displayed in one tab as line graphs, clinical notes appear as long-form text in another, transcripts are shown verbatim in a separate tab, and passive sensing data are visualized independently. While this design preserves modality-specific fidelity, it lacks integration across data types, placing the burden on clinicians to manually synthesize information across tabs when forming an overall understanding of the patient’s condition.}
    \label{fig:baseline-interface}
    \Description{}
\end{figure*}

\subsection{Study Design}
\label{sub:user-study:study-design}

\subsubsection{Conditions}
To understand if the computational narrative design of \systemName provided benefits for clinicians viewing multimodal patient data, we compare \systemName with a baseline interface \baselineName. 
Specifically, generated through the same computation pipeline as \systemName, \baselineName displays the information using the data collection dashboard design pattern, \ie laying out the four data categories in separate views (Fig.~\ref{fig:baseline-interface}). For instance, the view of passive sensing would only contain text descriptions and visualizations about the data facts extracted from passive sensing features. Moreover, we implemented \baselineName with a similar aesthetic to \systemName to minimize the impact of visual appearance on the study results.

We selected \baselineName over a traditional EHR as the baseline for two primary reasons. Traditional EHR systems might not integrate the full spectrum of data modalities we considered, and such a comparison might conflate the effect of the added data modalities with our narrative design. Additionally, we argue that \baselineName represents a typical near-future design (see \S\ref{subsec:related_work:dashboards-clinical-practice}) integrating multimodal data into existing EHR systems. By controlling the confounding variables between \systemName and \baselineName, we can focus on the effect of the narrative presentation compared to a typical dashboard baseline.

We paired the two conditions (\textit{\systemName}, \textit{\baselineName}) with two simulated patient scenarios (\textit{Lucy}, \textit{Gabriella}) that were similar in characteristics (\eg complexity, treatment methods). We leave the data for \textit{Alison} for demonstration and system introduction purposes.

\subsubsection{Measures}
We included the System Usability Scale (SUS)~\cite{brooke1996sus} and the NASA-TLX~\cite{hart1988development} to gauge and compare the usability and clinician's workload using both systems. 
Meanwhile, we designed a 10-question, 7-point Likert survey (see Appendix \S\ref{appendix:survey-questions-sys-eval}) used to capture quantitative user feedback regarding the quality and perception of both systems.
Specifically, four questions targeted \textbf{quality} of data insights, assessing the clinical relevance of the information, the system's ability to reveal hidden or complex insights, its effectiveness in supporting clinical decision-making, and the trustworthiness of the presented data.
The remaining six questions focused on user \textbf{perception}, evaluating aspects of system design such as the ease of interpreting the data, efficiency in reviewing patient information, integration of multimodal data, cohesiveness of the narrative presentation, time-saving potential, and support for effective communication during patient interactions.

\subsubsection{Procedures and Tasks}
Each study session comprised three stages: a) introduction to the study settings, b) two patient review tasks using \systemName and \baselineName with a counterbalanced order, and c) a post-study survey and a semi-structured interview to understand participants' experiences. Upon completion of the study, participants were thanked with a \$50 gift card.

\paragraph{Introduction}
We first provided an introduction to the study setting. We informed participants to imagine a hypothetical 4th encounter with a mental health patient whom they had last seen a month ago. We told them that their goal is to recap on the prior sessions using the data presented on the dashboards to recall the patient's conditions.

\paragraph{Patient Review Tasks}
We first gave a brief tutorial on the system (\textit{\systemName}, \textit{\baselineName}) to familiarize them with the interface.
Researchers then sent a link to direct participants to the target condition. We counterbalanced the interface condition to control order effect and rotated the patient scenarios (\textit{Lucy}, \textit{Gabriella}) to control the effect of the content on the conditions.
For each patient review, we gave participants 6 minutes with the target of giving a verbal summary of the patient's condition in the end. While the task is goal-oriented with a time limit, we informed participants that they would neither be graded on the number of insights found nor on the quality of the review to foster naturalistic review behavior. Besides, participants were encouraged to follow a think-aloud protocol to allow us to better understand their thinking process.

\paragraph{Post-study Survey and Interview}
After participants finished both conditions, participants filled out a survey where they reflected on the two systems' performance. At the end of each session, we conducted a semi-structured interview to collect qualitative feedback on \systemName. Building on their think-aloud feedback, we further asked them questions about their strategies for conducting patient review using both conditions, preference between the two conditions, attitudes towards the AI integration, and limitations and feedback regarding the functionality of \systemName. We also asked follow-up questions based on the participants' responses.

\subsection{Findings}
\label{sub:user-study:findings}

Since the study employed a within-subject design, we conducted quantitative analysis between \systemName and \baselineName using the Wilcoxon Signed-Rank test. Data processing and statistical analyses were conducted in R~\cite{r2013r}, using the \texttt{coin} and \texttt{rstatix} packages. We report the relevant statistical indicators, including the \p-value, \z-value, and the rank-biserial effect size (\eff). Additionally, we used the AI attitude score (1-10) and data familiarity (1-4, \ie number of reported data experience) as covariates and conducted regression analyses (using the \texttt{lmerTest} package) to investigate if they impact the evaluation results. 

Meanwhile, we analyzed the qualitative data through an iterative process of qualitative coding and affinity diagramming based on principles of thematic analysis~\cite{braun2006usinga, clarke2017thematic}. Two coders independently reviewed a subset of interview transcripts and notes to generate initial codes. Subsequently, in an iterative process, the two coders met for multiple rounds to discuss and refine the coding structure. Finally, the two coders collaboratively surfaced high-level themes from the codes to address our research questions.

\subsubsection{Quality: Expert Evaluation}

\begin{figure*}
    \centering
    \includegraphics[width=0.75\linewidth]{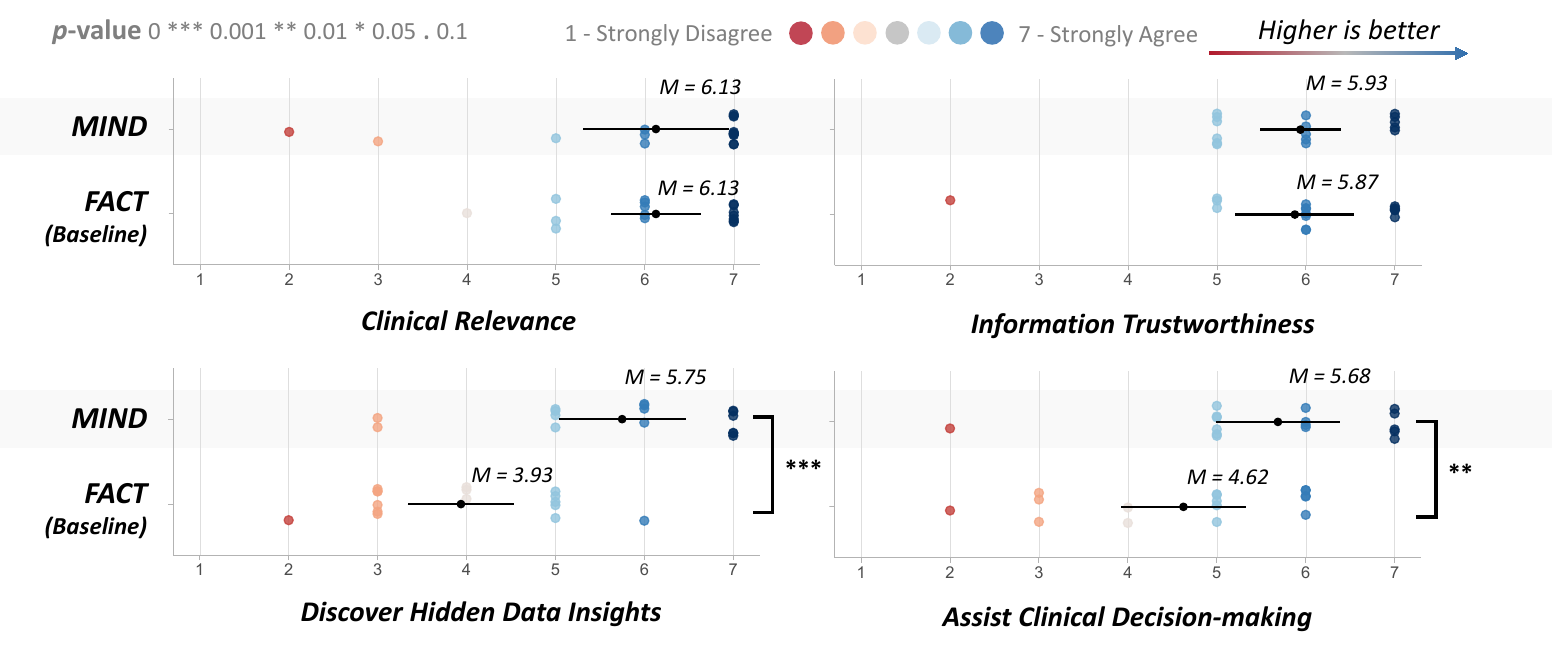}
    \caption{A comparison of information quality between \systemName and \baselineName using a 7-point Likert Scale. Each point in the chart represents a single record from a single participant. We map the values with a divergent color scheme from dark red (1-very low) to dark blue (7-very high). The black dot and line indicate the mean and 95\% confidence interval of the metric.}
    \Description{
    Results of the comparative evaluation between MIND and the baseline system FACT across four key dimensions: clinical relevance, information trustworthiness, ability to discover hidden data insights, and support for clinical decision-making. Each dot represents an individual clinician’s rating on a 7-point Likert scale, with higher values indicating stronger agreement. Mean scores with confidence intervals are displayed for both systems. While MIND and FACT were rated similarly for clinical relevance (M = 6.13) and information trustworthiness (MIND M = 5.93, FACT M = 5.87), MIND significantly outperformed FACT in its ability to help clinicians discover hidden data insights (MIND M = 5.75 vs. FACT M = 3.93, p < .001) and to assist clinical decision-making (MIND M = 5.68 vs. FACT M = 4.62, p < .01). These findings suggest that MIND’s narrative and integrative design improved clinicians’ ability to surface non-obvious patterns and apply insights directly to decision-making, while maintaining a comparable level of trustworthiness to the baseline.
    }
    \label{fig:result-quality}
\end{figure*}

Participants overall evaluated the information presented on \systemName as of good quality and significantly better than that presented on \baselineName (evaluation of DG4, Fig.~\ref{fig:result-quality}). Specifically, \systemName significantly outperforms \baselineName in providing quality information for clinicians to discover hidden data insights (\M{MIND}{5.75} \vs \M{FACT}{3.93}, \z=-3.39, \p<.001, \eff=.85), and in assisting clinical decision-making (\M{MIND}{5.68} \vs \M{FACT}{4.62}, \z=-2.94, \p=.004, \eff=.74). Meanwhile, the statistical test does not reveal differences regarding clinical relevance (\M{MIND}{6.13} \vs \M{FACT}{6.13}, \p=.438) and the trustworthiness of the information (\M{MIND}{5.93} \vs \M{FACT}{5.87}, \p=.922), with both high average scores suggesting that participants favors the information generated by our computation pipeline. An additional regression analysis using linear mixed effect models (R notation $\text{\textit{<Quality\_metrics>}} \sim \text{\textit{Condition}} + \text{\textit{AI\_Attitude}} + \text{\textit{Data\_Familiarity}} + \text{\textit{(1|PID)}}$, referent \baselineName condition) revealed no significance on participants' AI attitude and data familiarity, indicating the results are consistent across populations.

\subsubsection{Perception: System Design Evaluation}

\begin{figure*}[t]
    \centering
    \includegraphics[width=1\linewidth]{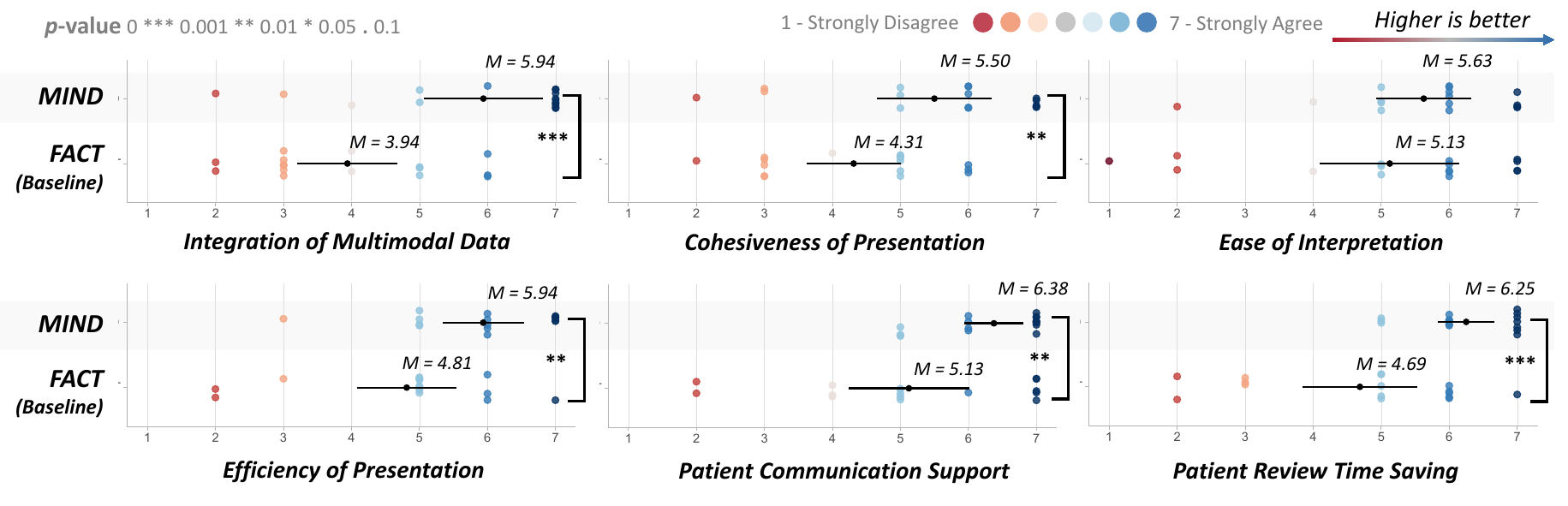}
    \caption{A comparison of the perception of system design between \systemName and \baselineName using a 7-point Likert Scale. Each point in the chart represents a single record from a single participant. We map the values with a divergent color scheme from dark red (1-very low) to dark blue (7-very high). The black dot and line indicate the mean and 95\% confidence interval of the metric.}
    \Description{
    Comparison of MIND and the baseline system FACT across six design-related dimensions measured on a 7-point Likert scale. Each point corresponds to one participant’s rating, with mean values and 95 percent confidence intervals shown in black. MIND was rated significantly higher than FACT on integration of multimodal data (MIND M = 5.94 vs. FACT M = 3.94, p < .001), cohesiveness of presentation (MIND M = 5.50 vs. FACT M = 4.31, p < .01), ease of interpretation (MIND M = 5.63 vs. FACT M = 5.13), efficiency of presentation (MIND M = 5.94 vs. FACT M = 4.81, p < .01), patient communication support (MIND M = 6.38 vs. FACT M = 5.13, p < .01), and patient review time saving (MIND M = 6.25 vs. FACT M = 4.69, p < .001). These results demonstrate that MIND’s narrative and integrative design improved both usability and communication outcomes, enabling clinicians to more efficiently synthesize multimodal information, convey insights to patients, and reduce review time compared to the baseline.
    }
    \label{fig:result-system-design}
\end{figure*}

Participants prefer \systemName as the interface for consuming multimodal data with significance in five out of the six measures tested (evaluation of DG1, 2; Fig.~\ref{fig:result-system-design}). \systemName significantly outperforms \baselineName in efficiency in reviewing patient information (\M{MIND}{5.94} \vs \M{FACT}{4.81}, \z=-2.78, \p=.006, \eff=.70), integration of multimodal data (\M{MIND}{5.94} \vs \M{FACT}{3.94}, \z=-3.20, \p<.001, \eff=.80), cohesiveness of the narrative presentation (\M{MIND}{5.50} \vs \M{FACT}{4.31}, \z=-2.90, \p=.004, \eff=.73), time-saving potential (\M{MIND}{6.25} \vs \M{FACT}{4.69}, \z=-3.20, \p<.001, \eff=.80), and effective communication during patient interactions (\M{MIND}{6.38} \vs \M{FACT}{5.13}, \z=-2.94, \p=.004, \eff=.73). Meanwhile, both \systemName and \baselineName showed positive ratings for ease of interpretation (\M{MIND}{5.63} \vs \M{FACT}{5.13}, \p=.180), though we reveal no significant difference comparing the two conditions. An additional regression analysis using linear mixed-effect models similar to the above test
revealed no significance on participants' AI attitude and data familiarity on their perception of the systems' design.

\subsubsection{Perception: System Usability Scale}

\begin{figure}[tbp]
    \centering
    \includegraphics[width=1\linewidth]{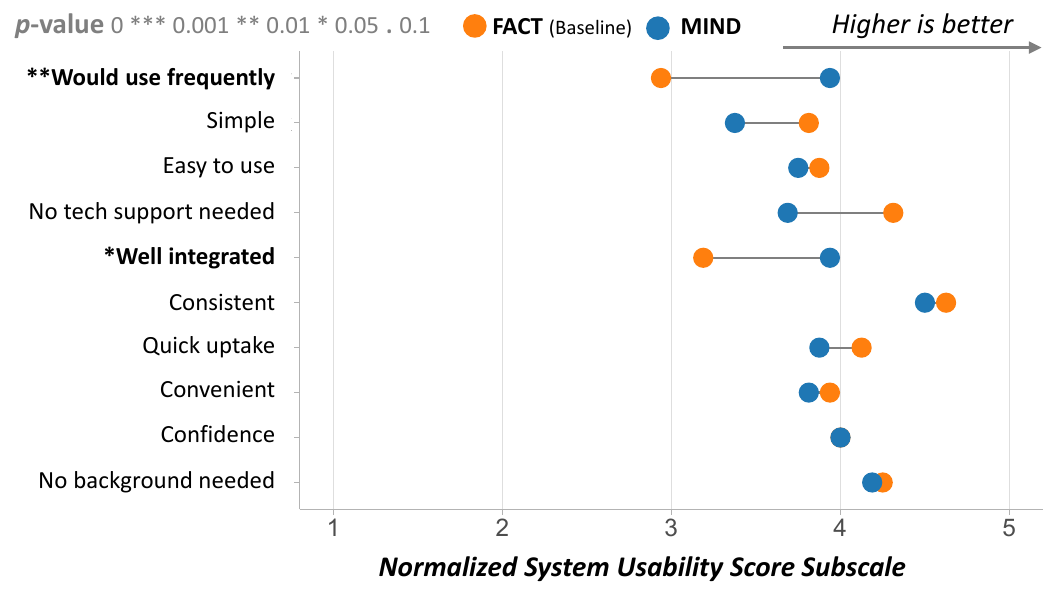}
    \caption{Normalized scores (higher is better) for the adapted SUS (\systemName \vs \baselineName). \textbf{Bolded items} indicate \p<0.05. }
    \Description{
    Normalized subscale scores from the adapted System Usability Scale (SUS), comparing MIND with the baseline system FACT. Higher scores indicate better usability. Across most subscales—including simplicity, ease of use, quick uptake, convenience, and user confidence—MIND and FACT performed similarly, with both systems rated above average. However, MIND was rated significantly higher on two items: clinicians reported that they would be more likely to use MIND frequently (p < .01) and that its functions felt more well integrated (p < .05). These results suggest that while both interfaces were generally usable, MIND’s narrative and integrative design features contributed to a stronger sense of cohesion and adoption potential in clinical workflows.
    }
    \label{fig:result-sus}
\end{figure}

Overall, \systemName showed a comparable system usability to \baselineName. %
A breakdown of the system usability subscales (Fig.~\ref{fig:result-sus}) revealed that participants perceived \systemName as a system they would more likely to use frequently (\M{MIND}{3.94} \vs \M{FACT}{2.94}, \z=-2.68, \p=.005, \eff=.67) and is better integrated (\M{MIND}{3.94} \vs \M{FACT}{3.19}, \z=-2.07, \p=.04, \eff=.52) than \baselineName. On the other hand, visual analysis revealed that clinicians preferred \baselineName over \systemName in a few aspects, especially its learnability (\ie no tech support needed \M{MIND}{3.69} \vs \M{FACT}{4.31}, \p=.125) and simplicity (\M{MIND}{3.38} \vs \M{FACT}{3.81}, \p=.219). However, the difference is not large enough for significance. %

\subsubsection{Perception: Workload}

Overall, we observed no significant difference across all measures (mental, physical, temporal, performance, effort, and frustration) of workload using the NASA-TLX scale (Fig.~\ref{fig:result-nasa-tlx}). 
We argue that the scores from NASA-TLX reflected more on the difficulty of the task itself, and merely integrating the information through narrative did not make the interpretation easier. Such scores suggested that clinicians maintained the agency in making clinical decisions using both \systemName and \baselineName.

\subsubsection{Perception: Clinicians' Feedback}

Qualitative data suggested that clinicians found \systemName effective for viewing multimodal patient data, though they also suggested several areas for improvement.

\paragraph{Clinicians prefer \systemName for its summarization capability, especially under time constraints.}
Across all 16 participants, clinicians preferred the \systemName dashboard over the \baselineName dashboard. \systemName was positively received for its summaries that distilled complex multimodal data into data insights. For example, P1 liked that ``\textit{it (\systemName) does summarize},'' whereas \baselineName was critiqued by having ``\textit{so much data to sort through}'' (P4). Under time constraints, clinicians commended \systemName because it provides a ``\textit{quick \& dirty snapshot}'' (P5) for them before entering a patient session.

\paragraph{Clinicians navigated \systemName using Session Recap and Timeline as key entry points.}
Clinicians (P3, P6, P9, P10, P13) identified the session recap and the timeline as two key entry points for finding insights in the dashboard. P9 commented that ``\textit{timeline on the bottom left was really the main way I navigated \systemName ... and the recap is helpful because it gives you that SOAP format summary.}'' P13 liked ``\textit{having the timeline visible}'' to ``\textit{orient}'' and guide ``\textit{what to do next}.'' P6 remarked that the timeline ``\textit{makes it easier to understand how things are progressing, instead of just snapshots}'' and that the timeline is ``\textit{closer to how I think about patient care}.'' Such feedback indicates that the narrative structure functioned effectively to provide clinicians with an intuitive interface for navigating multimodal data.

\paragraph{Verifiability and transparency are keys in fostering trust in \systemName}
During the interview session, we informed clinicians with a general description of the hybrid insight generation workflow. Clinicians overall commended the quality of the insights, saying that ``\textit{(The data insights) seem to be good summaries [of] what I would, in a bird's-eye view, want to know about how the month went [for the patient]}''. Additionally, when asked what designs helped them trust the system, they (P1, P2, P5, P14) emphasized that the affordance jumping to L2 (Drill-down panel) strengthens their confidence in the insights. P5 commented ``\textit{as long as I can check the raw data myself, I'll generally trust the AI}.'' Such a response further demonstrates how \systemName successfully embodies DG3. However, we also observed scenarios (P9, P16) where clinicians ``overtrust'' AI. For example, P9 states that ``\textit{I don't doubt AI ... I trust AI fully}.''

\paragraph{Challenges and concerns around using clinical transcripts in mental healthcare.}
Several clinicians raised privacy concerns and patients' discomfort about using verbatim session transcripts as a data source. Some (P1, P7) worried about the confidentiality of patients' data. P1 commented: ``\textit{The idea of having a transcript in somebody's medical record makes me nervous ... I'm not totally sure that the trade-off and, like, the [benefit and] risk on privacy [is worth it].}'' Meanwhile, P7 described that data breach would be ``\textit{really life changing for the patients}''. Others (P11, P16) questioned the practicality and validity of integrating clinical transcripts. For example, P16 thinks that transcripts are too wordy and overwhelming to be useful in daily practice, while P11 questions how the system would handle misalignment between patients' behavioral and linguistic patterns.

\paragraph{Room for Improvement}
While we made extensive efforts to prevent the dashboard from overwhelming clinicians, clinicians have differing opinions about \systemName's complexity. For example, P4 perceived \systemName as ``\textit{busy and crowded}'' while P8 reported ``\textit{too much text}'' on the screen. Meanwhile, both P10 and P14 reported excessing clicking interactions navigating the dashboard. In contrast, P7 believes that ``\textit{it's (P7's EHR system in real-world practice) not nearly as elegant as this (\systemName)}''. The disparity in the perception of \systemName's complexity surfaces future needs to personalize \systemName to suit different preferences among clinicians.

\paragraph{Functions beyond patient review}

We designed the Summary Today section as an open-ended summary that provides a hook for clinicians to communicate with patients after sessions. Clinicians (P6, P9, P11) appreciated its ability to streamline their workflow, while others (P3, P6, P8, P9) proposed imaginative uses for Summary Today. For instance, P8 reflected that ``\textit{if I could use this text generation (pointing at the Summary Today section) to write notes, that would be a really helpful tool ... if I could draft note at the end and kind of put these in, that would be helpful for me}.'' Future iterations of \systemName should include these functionalities to better incentivize clinicians to use the system.

\section{Discussion}
\label{sec:discussion}

Our work explored how a narrative dashboard can empower mental health clinicians in understanding multimodal data. While clinicians found data presented in both systems clinically relevant, \systemName's narrative design made multimodal data insights more useful in clinical practice. In this section, we reflect on our study and discuss how the narrative dashboard design could benefit broader stakeholders within and beyond mental healthcare. We also discuss the limitations of our research and envision future directions.

\subsection{Improving the Curation of Multimodal Data Insights}
In this study, we built a hybrid computation pipeline that transforms raw multimodal data into clinically relevant data insights. While our user study demonstrated that our pipeline can generate clinically relevant data insights, we only evaluated its performance on two simulated patient cases. A larger dataset containing heterogeneous real-world corpora to better evaluate the performance of the computation pipeline and, in turn, improve upon the current implementation. For example, research could attempt to design a full LLM-powered agentic workflow (\eg~\cite{choube2025gloss, abaskohi2025agentada}) or use multimodal models pre-trained with temporal passive sensing data (\eg~\cite{zhang2025sensorlm}). We advocate for future work to propose methods to collect clinician-aligned datasets related to multimodal data to better understand what counts as useful data narratives and improve automatic algorithms in curating multimodal data insights.

Additionally, our current implementation does not explicitly account for the noise inherent within each data modality. For instance, passive sensing data can be inconsistent due to sensor variability, while clinical and active sensing data are subject to noise from imperfect recall and subjective survey scoring~\cite{adler2024detectiona}. Although our multimodal approach can partially counteract noise by revealing cross-modal consistencies and inconsistencies, the system does not formally quantify uncertainty—neither in the source data nor in the final curated insights. Consequently, judging the validity and weight of these insights remains a task for the clinician, with \systemName serving a supplementary role for synthesizing a data-driven narrative~\cite{nghiem2023understandingc}. Future work could therefore develop methods to better quantify the noise in the data and integrate explicit uncertainty cues, whether quantitative (\eg confidence scores~\cite{zhang2020effect}) or qualitative, to help clinicians effectively interpret the derived insights and ensure they are ``actionable'' for clinical decision-making~\cite{adler2024detectiona}.

\subsection{Towards Tailored and Shared Decision- making between Clinician and Patient}

\systemName includes a Patient Communication section, which presents a potential portal for linking data narratives with patients. However, our tool 1) remains primarily designed for clinicians and the clinical workflow, and 2) represents a design for ``common grounds'' of mental health practice instead of tailoring for a specific mental health challenge. Future work should therefore seek to adapt \systemName to support both clinicians' workflow and the patient-clinician collaborative decision-making process during clinical sessions~\cite{mentis2017craftinga}, and afford clinicians and patients tailor the interface to their specific conditions, measurements, data modalities (\eg patient journals), and treatment plans. One promising approach is to use malleable interfaces, for example, morphing narrative dashboard content into data comics or infographics~\cite{wang2019comparing}, to better communicate patient/treatment-specific data insights to patients (\eg~\cite{stromel2024narrating}). By involving patients in the process, we could foster a greater willingness to share passive sensing data and take active sensing surveys, ultimately creating a positive feedback loop between enhanced data sharing and more data-driven clinical sessions. Furthermore, achieving a shared understanding~\cite{MCCABE2021114208} of multimodal data could promote continuity of care--for instance, by enabling just-in-time notifications based on sensing data to advocate health behaviors and improve treatment adherence~\cite{sharmin2015visualizationb}.

\subsection{Narrative Dashboards beyond Mental Health}
Our study demonstrated the potential of using narrative dashboards as a means to communicate multimodal data insights to mental health clinicians. We envision that similar narrative dashboards could be extended to domains beyond mental healthcare as a supplement to their current workflow. For instance, cardiologists and oncologists similarly face the challenge of integrating multimodal data sources (\eg imaging, subjective patient reports, clinical notes) in a short amount of time to inform critical diagnostic and treatment decisions. Future work could seek to design and evaluate domain-specific narrative dashboards as a way to address the unique challenges in integrating multimodal data in other healthcare domains, with the ultimate goal to improve clinical outcomes.

\subsection{Ethical Considerations Applying LLM-powered Systems in Health}

In addition to the potential to further improve and expand \systemName, the integration of an LLM/rule-based hybrid pipeline into \systemName raises broader ethical considerations for generative technologies in clinical workflows. For instance, recent works revealed risks related to bias, sycophancy, and stigmatizing language when applying LLMs in mental health contexts~\cite{moore2025expressing}, which can cause serious ramifications~\cite{booth2025teen}. While we designed \systemName to be clinician-facing, mitigating the direct risk of exposing an underdeveloped algorithm to users, it requires sustained future efforts to understand how computationally curated data insights impact clinicians' decision-making process despite its perceived usefulness. Specifically, content curated by LLMs may be \textit{biased} towards dominant or easily accessible data points, potentially creating a ``generative echo chamber''~\cite{sharma2024generative} that restricts the breadth of information available to the clinician.

A closely related concern is the \textit{trust} in LLM-powered systems. In our design, \systemName employed multiple methods (\eg showing data source types, supporting drill-down fact-checking) to foster adequate trust towards the data insights. While evaluation demonstrated most clinicians actively check the raw data before making their own judgments, some are more inclined to ``overtrust'' AI after only a few successes with the generated data insights. Such over-reliance raises concerns regarding the \textit{accountability} of human clinicians working with AI and, without mitigation, may ultimately hamper treatment quality and compromise patient safety.

Additionally, \textit{privacy} is also a concern reflected through our study. Some clinicians expressed reservations about using raw clinical transcripts as a data modality, recognizing that these sensitive records can also serve as potential legal evidence that may put both the patient and the clinician at risk. Subsequently, we advocate for future research to design privacy-preserving data infrastructures to capitalize on their benefits while mitigating inherent limitations.

\subsection{Limitations and Future Work}
There are several limitations in our study that inform future work opportunities. First, we used homogeneous simulated patient cases in our user study (\eg SOAP notes, depression-only patients) to treat it as a controlled variable. However, real-world cases could be more heterogeneous (\eg missing data, diverse note styles), and the quality of the generated insights may vary due to those variances.

Second, we recruited our participants through a combination of convenience, purposive, and snowball sampling, which may not be representative of the demographics of mental health clinicians. For instance, our sample likely over-represented clinicians who are more willing to experiment with new technologies. This is consistent with our population's overall positive attitude towards AI, which might be a contributing factor to why clinicians appreciated the computationally generated data insights.

Third, the evaluation of our computation pipeline might be limited to specific use cases. Although two clinicians reviewed and approved the generated content for the user study, this internal validation may be subjective and does not fully account for challenges using LLMs, such as potential biases during the narrative curation process. Future work should therefore combine expert assessments with large-scale benchmarking to evaluate the pipeline's performance across diverse scenarios and mitigate model risks.

Finally, we conducted our experiments using a controlled lab study. While minimizing confounding variables, it did not fully capture the nuances of actual practice (\eg strict time constraints, interruptions). Synthesizing the three limitations above, future work should seek to further diversify the patient data used for evaluation, expand participant demographics through stratified sampling, and observe \systemName's use in real-world practice (\eg through contextual inquiry~\cite{karen1993contextual}, especially how it might complement current EHR systems) to better understand its applicability and potential for integration into clinical workflows.

\section{Conclusion}
\label{sec:conclusion}

In this paper, we propose \systemName, a multimodal data-integrated narrative dashboard designed to empower mental health clinicians for patient review. Through a co-design study with five mental health experts, we present a data-driven narrative with multimodal data insights based on a hybrid computational pipeline.
A user study with 16 licensed clinicians shows that \systemName fulfilled our design goals, demonstrating its ability to efficiently reveal hidden, clinically relevant insights and its potential to support clinical decision-making. Clinicians found the narrative structure intuitive and valued the system's emphasis on verifiability through the drill-down panel. \systemName demonstrated the feasibility and power of combining careful computation curation with narrative dashboards in presenting multimodal data. We further advocate for future work towards real-world longitudinal evaluations of \systemName in clinical practice, as well as applying narrative dashboards towards other healthcare domains and diverse healthcare stakeholders.

\section*{Acknowledgment of the Use of AI}
We acknowledge the use of Generative AI tools to improve the grammar, style, and readability of this manuscript. These tools were used in our computation pipeline (see details in \S\ref{sec:computation}) and text editing, but played no role in the data analysis, interpretation, or generation of the core findings presented.

\begin{acks}
Research reported in this publication was supported in part by the National Cancer Institute of the National Institutes of Health under Award Number R01CA301579. The content is solely the responsibility of the authors and does not necessarily represent the official views of the National Institutes of Health. We also thank the Columbia University Research Stabilization Grant and the Northeastern University Tier-1 Research Seed Grant for supporting this research project. We want to thank all participants for taking part in our study. Additionally, we extend our thanks to the anonymous reviewers for their constructive feedback.
\end{acks}
\balance

\bibliographystyle{ACM-Ref-Format}
\bibliography{
bib/others,
bib/zotero_export,
bib/Research-Knowledge,
bib/zotero_export_new
}
\clearpage

\appendix
\onecolumn

\section{Artifacts Used in Co-Design Studies}
\label{appendix:artifacts-codesign}

\begin{figure*}[htb!]
    \centering
    \includegraphics[width=1\linewidth]{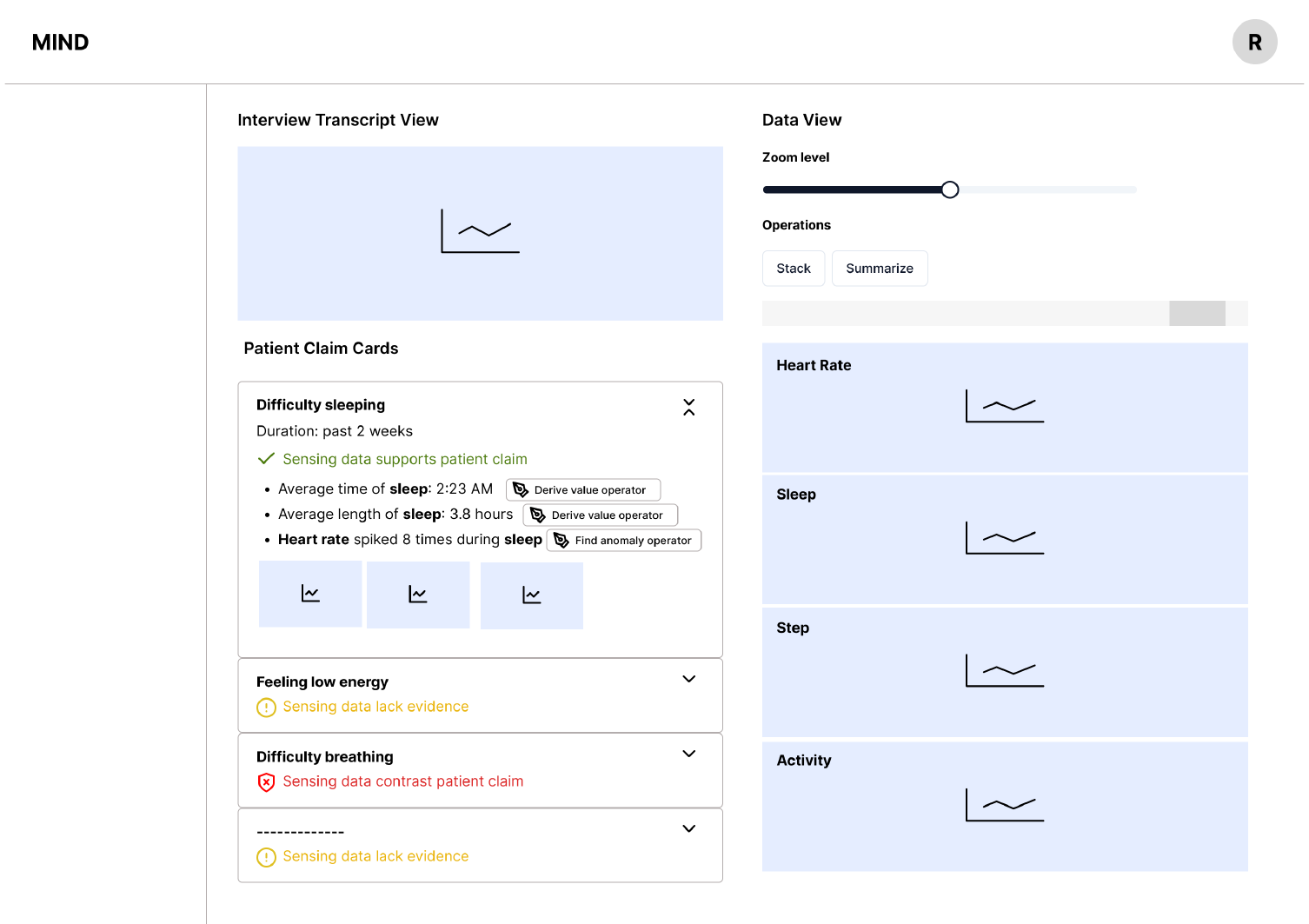}
    \caption{First low-fidelity prototype (Fig.~\ref{fig:iteration-progress}.C1). Clinicians pointed out problems such as vague wording, excessive use of graphics, charts being too small within the cards, and an overwhelming dashboard with limited structure.}%
    \Description{
    The first low-fidelity prototype of MIND, corresponding to the early design concept of stacked text–chart cards (see Fig. 2C1). The interface organized patient information into discrete “claim cards” that linked sensing data to reported symptoms, while providing transcript and raw data views on the side. Although this version demonstrated the potential of aligning patient-reported claims with sensor-derived evidence, clinicians identified several usability issues. These included vague or overly technical wording in the claim cards, excessive reliance on small embedded graphics that were difficult to read, and an overall dashboard layout that felt visually cluttered and insufficiently structured. Feedback from this iteration highlighted the need to streamline visual presentation, clarify language, and provide stronger organizational scaffolding to support clinician interpretation.
    }
    \label{fig:appendix:system-initial}
\end{figure*}

\begin{figure*}[htb!]
    \centering
    \includegraphics[width=1\linewidth]{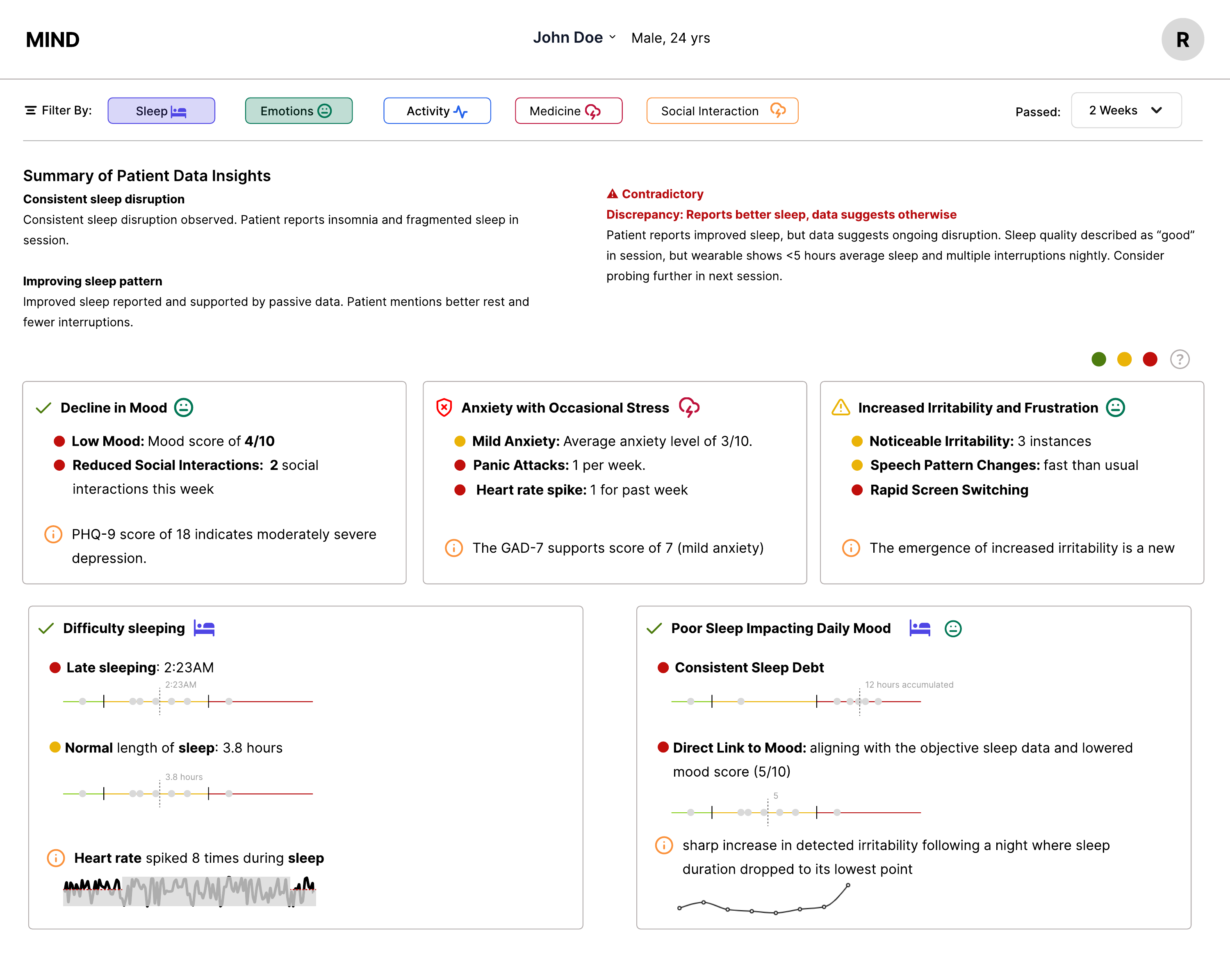}
    \caption{A representative wireframe of an earlier design artifact of \systemName (Fig.~\ref{fig:iteration-progress}.C2). At this stage, the design focuses on presenting scattered ``data facts'' to clinicians without a clear narrative structure.}
    \Description{
    A mid-fidelity wireframe prototype of MIND, corresponding to the second design concept (Fig. 2C2). In this version, the system organized patient insights into a grid of “data fact” cards that summarized specific observations from clinical and sensing data, such as mood decline, sleep disruption, or anxiety episodes. The cards integrated icons, bullet points, and short textual descriptors to communicate each observation, with filtering options allowing clinicians to view facts by domain (e.g., sleep, emotions, activity, medication, or social interaction). While this design made multimodal data accessible at a glance, clinicians reported that the dashboard felt fragmented and lacked a coherent narrative structure. The scattered cards required additional effort to synthesize into a clinical storyline, and contradictory information across cards created uncertainty without guidance on interpretation. Feedback from this stage emphasized the need to move beyond presenting isolated facts and toward a narrative-driven format that could better align with clinical reasoning practices.
    }
    \label{fig:appendix:system-pre-nar}
\end{figure*}

\begin{figure*}[t]
    \centering
    \includegraphics[width=0.86\linewidth]{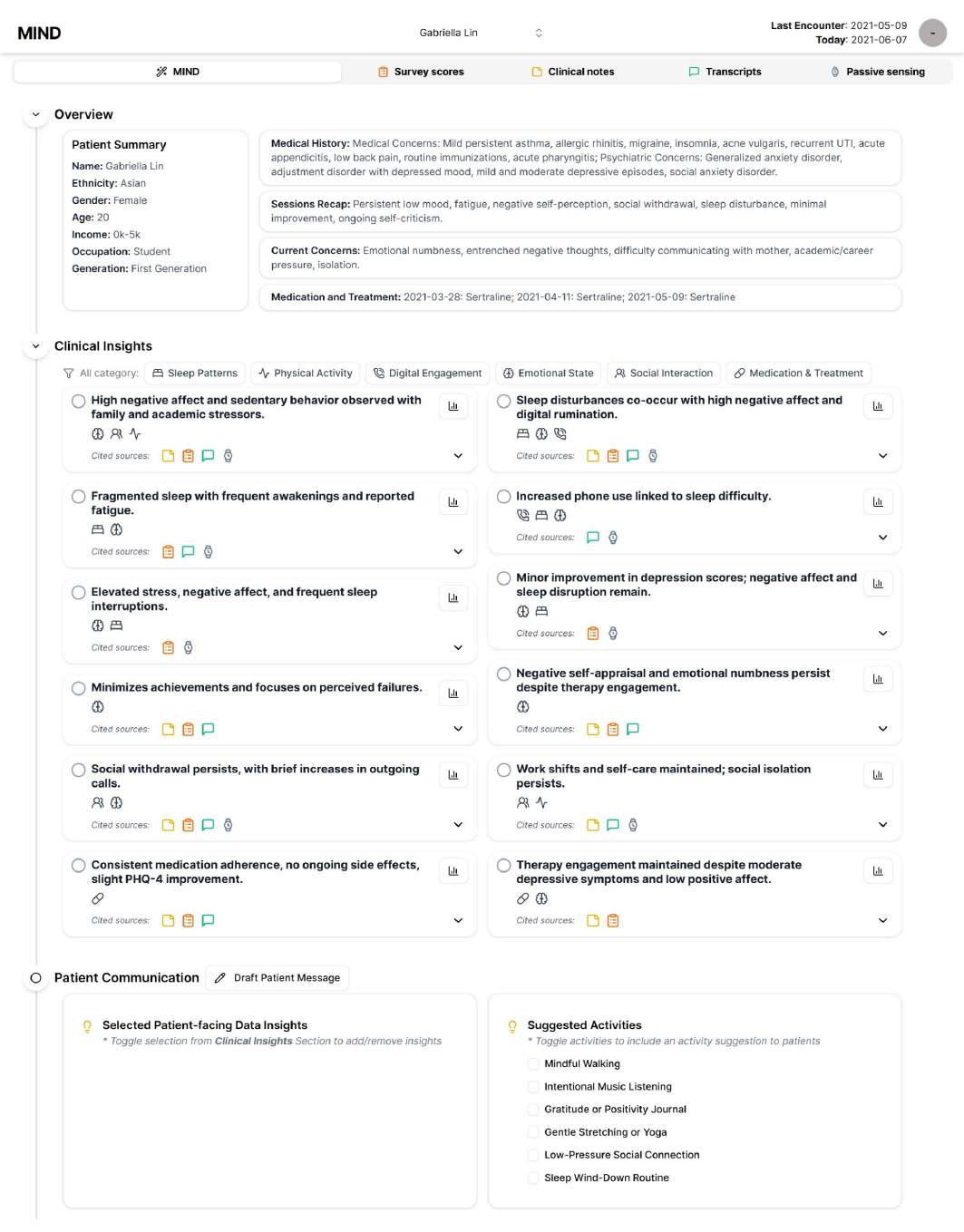}
    \caption{The first interactive version of \systemName (Fig.~\ref{fig:iteration-progress}.C3, scrolling content expanded). At this stage, \systemName incorporates the three-act narrative structure, but without a clear timeline structure or narrative inside the \textit{Clinical Insights} section.}
    \Description{
    The first interactive prototype of MIND, corresponding to the third design concept (Fig. 2C3). This version introduced a more structured three-act narrative organization, with distinct sections for an overview of patient background, detailed clinical insights, and a patient communication panel for generating draft messages. The Clinical Insights section presented a scrollable set of modular insight cards, each describing a specific observation such as sleep disturbances, therapy engagement, or stress levels, accompanied by cited data sources and an expandable drill-down button. Suggested activities could be automatically drafted into patient-facing messages by selecting relevant insights. While this version marked progress toward integrating narrative elements and supporting clinician–patient communication, participants noted that the lack of an explicit timeline structure made it difficult to situate insights within the patient’s broader trajectory. The prototype highlighted the potential of structured storytelling but underscored the need for temporal organization to better align with clinical reasoning and review workflows.
    }
    \label{fig:appendix:system-after-nar}
\end{figure*}

We provide three checkpoint designs (corresponding to Fig.~\ref{fig:iteration-progress}.C1-C3) of \systemName (Fig.~\ref{fig:appendix:system-initial}, Fig.~\ref{fig:appendix:system-pre-nar}, and Fig.~\ref{fig:appendix:system-after-nar}) that mark significant design changes to demonstrate how \systemName improved throughout the co-design process.

\newpage

\newpage

\section{Data Features}
\subsection{Survey Scores}
\label{appendix:ps-survey}

We document the survey scores used for the mock patients in Tab.~\ref{tab:survey-features}.

\begin{table*}[ht]
\small
\caption{Selected Routine Survey Features from GLOBEM~\cite{xu2022globema}}
\Description{
Selected routine survey features used in MIND, drawn from established mental health instruments included in the GLOBEM dataset. The PHQ-4 provides a brief 4-item screen for anxiety and depression, with total scores ranging from 0 to 12 and cutoffs indicating normal, mild, moderate, and severe symptom levels. Its subscales capture anxiety (equivalent to the GAD-2) and depression (equivalent to the PHQ-2), each scored from 0 to 6. The PSS-4 is a 4-item measure of perceived stress with scores from 0 to 16, where higher scores reflect greater stress. Positive Affect and Negative Affect are each measured by 5-item subscales of the PANAS-SF, scored from 5 to 25, capturing self-reported affective states. Together, these surveys provide standardized, validated self-report data that complement sensing and clinical sources, enabling MIND to incorporate subjective measures of mood, stress, and affect into its multimodal insight generation.
}
\centering
\begin{tabularx}{\linewidth}{l X}
\toprule
\textbf{Feature Name} & \textbf{Description} \\
\midrule
PHQ-4~\cite{lowe20104item} & A 4-item survey for mental health conditions, including anxiety and depression. Each item is rated 0-3, with a total score range of 0-12, calculated by summing all four items. Interpretation of total scores: Normal (0-2), Mild (3-5), Moderate (6-8), Severe (9-12).\\

PHQ-4 Anxiety~\cite{lowe20104item} & A subscale of PHQ-4 (equivalent to GAD-2~\cite{kroenke2007anxiety}), consisting of the first two items of PHQ-4. Scores range from 0-6, with scores greater than 3 suggesting anxiety.\\
PHQ-4 Depression~\cite{lowe20104item} &  A subscale of PHQ-4 (equivalent to PHQ-2~\cite{kroenke2003patient}), consisting of the last two items of PHQ-4. Scores range from 0-6, with scores greater than 3 suggesting depression.\\

PSS-4~\cite{cohen1983global} &  A 4-item measure of perceived stress. Each item is rated 0-4, with a total score range of 0-16. Higher scores indicate greater levels of perceived stress, with scores greater than 6 suggesting high stress.\\

Positive Affect~\cite{watson1988development} &  A 5-item subscale of the Positive and Negative Affect Schedule Short Form (PANAS-SF) that measures positive affect. Each item is rated 1-5, with a total score range of 5-25. Higher scores reflect greater positive affect.\\
Negative Affect~\cite{watson1988development} &  A 5-item subscale of the Positive and Negative Affect Schedule Short Form (PANAS-SF) that measures negative affect. Each item is rated 1-5, with a total score range of 5-25. Higher scores reflect greater negative affect.\\
\bottomrule
\end{tabularx}
\label{tab:survey-features}
\end{table*}

\subsection{Passive Sensing Data}
\label{appendix:ps-data}
We document the passive sensing data features used for the mock patients in Tab.~\ref{tab:passive-features}.

\begin{table*}[ht]
\small
\caption{Selected Daily Passive Sensing Features from GLOBEM~\cite{xu2022globema}}
\Description{
Selected daily passive sensing features used in MIND, adapted from the GLOBEM dataset. These features capture behavioral and physiological markers derived from mobile and wearable sensors. Location-based features include distance traveled (converted from meters to miles), time spent at home (minutes to hours), and routine consistency (scaled from 0–1 to 0–100, indicating irregular to regular patterns). Screen-related features track the number of phone unlock episodes and total screen time in hours. Sleep-related features include awakening episodes, total sleep duration, and estimated bedtime and wake time. Step-based features capture sedentary episodes (inactive periods) and total step count. Together, these features provide continuous, unobtrusive measures of daily activity, sleep, mobility, and digital behaviors, offering complementary context to survey and clinical data in characterizing patients’ lived experiences.
}
\centering
\begin{tabularx}{\linewidth}{l l l X}
\toprule
\textbf{Feature Name} & \textbf{Orginal Unit} & \textbf{Translated Unit} & \textbf{Description} \\
\midrule
\multicolumn{4}{l}{\textbf{\underline{Location}}} \\
Distance Traveled & meters & miles & Total distance traveled. \\
Time Spent at Home & minutes & hours & Time spent at home. \\
Routine Consistency & 0-1 & 0-100 & A score between 0-100: 0 denotes irregular and 100 regular. \\
\midrule
\multicolumn{4}{l}{\textbf{\underline{Screen}}} \\
Phone Unlocks & count & times & Number of unlock episodes. \\
Total Screen Time & minutes & hours & Total duration of all unlock episodes. \\
\midrule
\multicolumn{4}{l}{\textbf{\underline{Sleep}}} \\
Awakening Episodes & count & times & The number of times a person wakes up during sleep. \\
Total Sleep & minutes & hours & Total sleep duration. \\
Bedtime & minutes & hours & Time of first bedtime after midnight. \\
Wake Time & minutes & hours & Time of last wake time after midnight. \\
\midrule
\multicolumn{4}{l}{\textbf{\underline{Steps}}} \\
Inactive Periods & count & times & Number of sedentary episodes (periods of inactivity).\\
Total Steps & count & count & Total step count. \\
\bottomrule
\end{tabularx}
\label{tab:passive-features}
\end{table*}

\section{Simulated Patient Cases}
\label{appendix:synthetic-data-generation}

We document patient-seed interview link pair in Tab.~\ref{tab:synthetic-patients}.

\begin{table*}[!h]
\small
\caption{Simulated Patient Cases}
\Description{
Simulated patient cases used in the study, drawn from the GLOBEM dataset. Three mock patients—Gabriella Lin (P# 963), Lucy Sutton (P# 1044), and Alison Daniels (P# 1077)—were selected as representative cases. Each simulated patient included multimodal data streams (survey responses, passive sensing, and transcripts) and was anchored by a seed interview session, linked here via publicly available video recordings. These simulated cases allowed for consistent evaluation of MIND across participants while preserving ecological realism through the use of authentic multimodal data grounded in standardized patient interviews.
}
\centering
\begin{tabular}{lll} %
\toprule
\textbf{Simulated Patient Name} & \textbf{P\# in GLOBEM} & \textbf{Seed Interview Session Link} \\
\midrule
Gabriella Lin  & 963  & \url{https://www.youtube.com/watch?v=JKUFWK6iSsw} \\
Lucy Sutton    & 1044 & \url{https://www.youtube.com/watch?v=7LD8iC4NqXM} \\
Alison Daniels & 1077 & \url{https://www.youtube.com/watch?v=4YhpWZCdiZc} \\
\bottomrule
\end{tabular}
\label{tab:synthetic-patients}
\end{table*}

\section{Survey Questions for System Evaluation}
\label{appendix:survey-questions-sys-eval}

\noindent \textbf{Quality}
\begin{enumerate}
   \item \textbf{Clinical Relevance}: The information presented in the system was clinically relevant.
     \item \textbf{Information Trustworthiness}: I felt I could trust the information provided by the system.
    \item \textbf{Discover Hidden Data Insights}: The system's presentation of data helped me identify important relationships or patterns that were not immediately obvious.
    \item \textbf{Assist Clinical Decision-making}: Reviewing the patient's data in the system increased my confidence in my clinical assessment.
\end{enumerate}

\noindent \textbf{Perception}
\begin{enumerate}
\setcounter{enumi}{4}
    \item \textbf{Integration of Multimodal Data}: It was easy to bring together information from different data sources to understand the overall patient situation.
    \item \textbf{Cohesiveness of Presentation}: The system presented the patient's health information as a clear, cohesive story rather than as separate, unrelated facts.
    \item \textbf{Ease of Interpretation}: I could easily interpret the data presented by the system.
    \item \textbf{Efficiency of Presentation}: Using this system made reviewing patient information more efficient.
    \item \textbf{Patient Communication Support}: Using this system could help facilitate more effective conversations during patient sessions.
    \item \textbf{Patient Review Time Saving}: Using this system could help me save time when reviewing a patient's condition.

\end{enumerate}

\section{Supplemental Evaluation Results}

\begin{figure*}[h]
    \centering
    \includegraphics[width=1\linewidth]{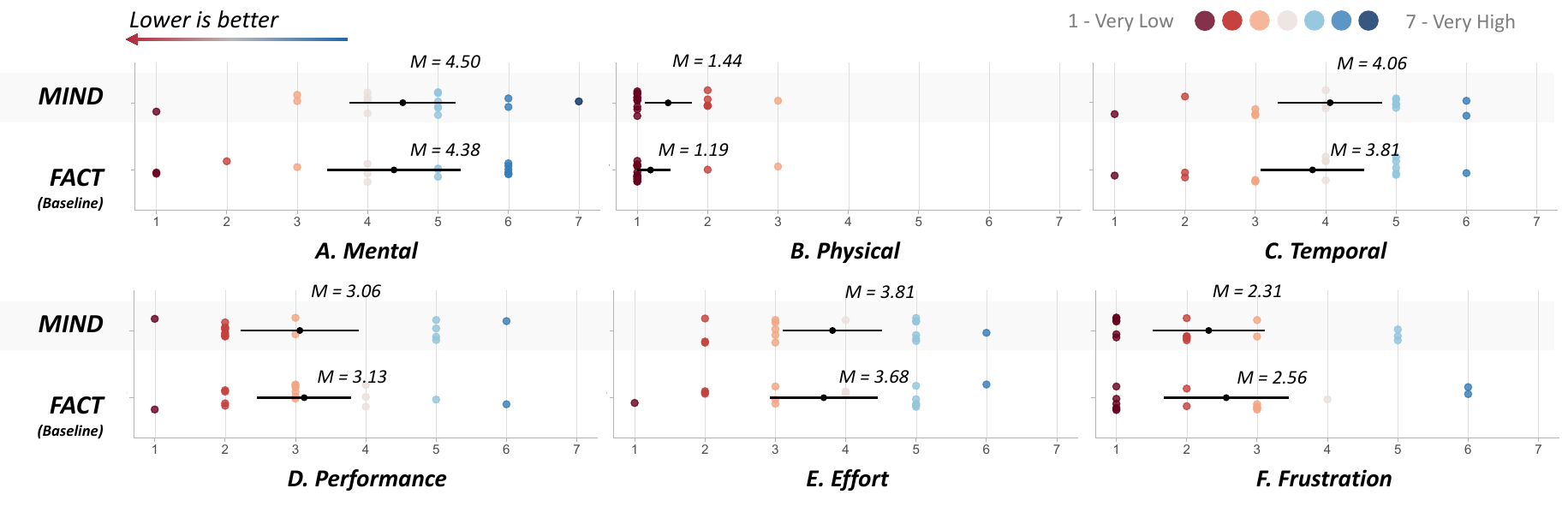}
    \caption{A comparison of perceived workload between \systemName and \baselineName on NASA-TLX. Each point in the chart represents one individual record from one participant. We map the values with a divergent color scheme from dark red (1-very low) to dark blue (7-very high). The black dot and the black line indicate the mean and 95\% confidence interval of the metric.}
    \Description{
    Perceived workload comparison between MIND and the baseline system FACT, measured using the NASA-TLX across six dimensions: mental demand, physical demand, temporal demand, performance, effort, and frustration. Each point represents an individual participant’s rating on a 7-point scale, with lower scores indicating reduced workload. Mean values and 95 percent confidence intervals are shown with black dots and lines. Across most dimensions, MIND and FACT received comparable ratings: for example, mental demand (MIND M = 4.50, FACT M = 4.38) and performance (MIND M = 3.06, FACT M = 3.13). Small differences were observed in physical demand (MIND M = 1.44 vs. FACT M = 1.19), temporal demand (MIND M = 4.06 vs. FACT M = 3.81), effort (MIND M = 3.81 vs. FACT M = 3.68), and frustration (MIND M = 2.31 vs. FACT M = 2.56). Overall, results suggest that MIND did not introduce additional workload compared to the baseline, maintaining comparable levels of cognitive, temporal, and emotional demands despite presenting more structured and synthesized information.
    }
    \label{fig:result-nasa-tlx}
\end{figure*}

We show the NASA-TLX results in Fig.~\ref{fig:result-nasa-tlx}.

\end{document}